\documentclass[prb,preprint]{revtex4-1}

\usepackage{graphicx}
\usepackage{appendix}
\usepackage[]{algorithm2e}
\usepackage{bm}
\usepackage{amsmath}
\usepackage[T1]{fontenc}
\usepackage{setspace}

\begin{document}

\title{Weyl's problem: A computational approach}

\author{Isaac Bowser}
\email{isaac\_bowser@taylor.edu}
\affiliation{Physics and Engineering Department, Taylor University,
  236 West Reade Ave., Upland, Indiana 46989}

\author{Joshua Kiers}
\email{jokiers@live.unc.edu}
\affiliation{Department of Mathematics, University of North Carolina at Chapel Hill, CB\# 3250
  Phillips Hall, Chapel Hill, North Carolina 27599-3250}
  
\author{Ken Kiers}
\email{knkiers@taylor.edu}
\affiliation{Physics and Engineering Department, Taylor University,
  236 West Reade Ave., Upland, Indiana 46989}

\author{Erica Mitchell}
\email{erica\_mitchell@taylor.edu}
\affiliation{Physics and Engineering Department, Taylor University,
  236 West Reade Ave., Upland, Indiana 46989}

\date{\today}

\begin{abstract}
The distribution of eigenvalues of the
  wave equation in a bounded domain is known as Weyl's problem.  We describe
  several computational projects related to the cumulative state
  number, defined as the number of states having wavenumber up to a
   maximum value.  This quantity and its derivative, the
  density of states, have important applications in nuclear physics,
  degenerate Fermi gases, blackbody radiation, Bose-Einstein
  condensation and the Casimir effect.  Weyl's theorem states that, in
  the limit of large wavenumbers, the cumulative state number depends
  only on the volume of the bounding domain and not on its shape.
  Corrections to this behavior are well known and depend on the
  surface area of the bounding domain, its curvature and other
  features.  We describe several  projects that allow
  readers to investigate this dependence for three  bounding
  domains -- a rectangular box, a sphere, and a circular cylinder.
  Quasi-one- and two-dimensional systems can be analyzed by
  considering various limits.   The projects   have 
  applications in statistical mechanics, but can also be integrated
  into quantum mechanics, nuclear physics, or computational physics
  courses.
\end{abstract}

\maketitle

\section{\label{sec:intro} Introduction}

From the late 1800s to the early 1970s, Weyl and many others studied
the distribution of eigenvalues of the wave equation,
\begin{equation}
  -\nabla^2\psi = k^2\psi \,,
  \label{eq:wave-equation}
\end{equation}
inside a finite domain such as a sphere, cube, or other shapes.  Many
 physical systems are described by Eq.~\eqref{eq:wave-equation},
differing only in the definition of the wavenumber $k$ and in the
boundary conditions.  The cumulative state number (or
mode number or integrated density of states), $N(k)$,
is the number of distinct states at or below some
value of $k$; the derivative of $N(k)$  is the density of
states.  Knowledge of the functional form of $N(k)$ and/or the density
of states is of great importance in studies of blackbody radiation,
the Casimir effect, acoustics and elasticity, Bose-Einstein
condensation, nuclear physics and degenerate Fermi gases.
Reference~\onlinecite{baltes-hilf} provides an excellent overview of this
field, including many  details and an extensive review of the
literature up to the early 1970s.

According to Ref.~\onlinecite{baltes-hilf}, Pockels\cite{pockels} was the
first to count the eigenvalues of Eq.~\eqref{eq:wave-equation},
with subsequent and related work done by Rayleigh, Lorentz, Reudler,
Weyl, and others.~\cite{baltes-hilf,reudler,weyl}  Of particular
importance is the  large $k$ behavior of
$N(k)$, which was shown to be independent of the   shape of the bounding domain and to depend only on its 
  volume, a result now known as Weyl's theorem.\cite{weyl}
Corrections to the infinite volume result depend on the surface area of the bounding
domain,  its curvature, and on other geometrical features. There is
an extensive   literature on  the importance
of surface and shape effects in various areas of physics (see
Ref.~\onlinecite{baltes-hilf}, p.~11).

In this paper we study Weyl's problem in the context of  a rectangular parallelepiped, a sphere
and a circular cylinder, which were  chosen because the
eigenvalues are well known and may be computed   straightforwardly.  For the rectangular and
cylindrical cases we can analyze quasi-one- and two-dimensional
structures, which are of  theoretical and experimental interest.

Our focus  is on comparing exact (discrete) results for
$N(k)$ with the approximate asymptotic result.  
Historically, there has been an interesting interplay between
computational and analytical  approaches to the
problem.\cite{footnote1}

For the bounding
domains that we consider, the eigenstates may be labeled by triplets
of integers, and the states contributing to $N(k)$ for a particular
value of $k$ may be represented by points on a three-dimensional
lattice.  These points are bounded by a surface in this space, and
calculation of $N(k)$  reduces to summing  the points within
this bounding surface.  Students will likely be familiar with  
the cubic bounding domain, for which the bounding surface in the complementary
space is one eighth of a sphere.  For a general rectangular domain,
the bounding surface becomes one eighth of an ellipsoid.

For large $k$ it is appropriate to treat $k$ as a
continuous variable and to replace sums over points on a lattice by
integrals.  Textbook derivations of $N(k)$ and related quantities
usually follow this approach and take the bounding domain to be a
cube.  In this work we show how we can follow this  approach for
all three of the bounding domains we consider, and  derive both the
leading volume term in the asymptotic expansion as well as the
(subleading) surface area terms.  The spherical and cylindrical cases
are more involved and are treated in the appendices.

This paper includes several computational projects to help students
verify and analyze various topics related to Weyl's problem.  We
include algorithms and pseudocode that encourage students to approach
the problem from a geometrical perspective.  Although as physicists we
are generally adept at switching between sums and integrals, our
students may perceive this as a sleight of hand.  The computations we
describe will improve students' computational abilities, while
simultaneously helping them to develop their intuition and ability to
understand the limitations of the simplifying assumptions.  For
example, one might wonder about the applicability of the continuum
approximation when the bounding domain is quasi-one- or
two-dimensional.  In this case, the wavenumber is much more sensitive
to the quantum numbers associated with the small dimensions, and it is
difficult to imagine that approximating a sum by an integral is
justified in this case.\cite{fisher} We shall see that the approach to
asymptopia is very slow for systems that are quasi-one- or
two-dimensional.  And yet such systems do eventually approach the
asymptotic limiting behavior required by Weyl's theorem.

Although our  focus is almost
exclusively on Weyl's problem, the literature   gives many
examples of possible applications.  Molina\cite{molina} and
Guti\'{e}rrez and Y\'{a}\~{n}ez\cite{gutierrez-yanez} consider
thermodynamic quantities related to an ideal gas in a finite-sized
container, focusing on effects due to the shape of the container, such
as the ratio of the surface area to the volume.
References~\onlinecite{pathria} and \onlinecite{bereta-et-al} analyze
Bose-Einstein condensation in a finite-sized one-dimensional box and
in spherically symmetric traps, respectively.  The latter includes a
discussion of a quasi-two-dimensional ``bubble trap,'' as well as
references to relevant experimental work in the context of cold gases.\cite{bereta-et-al}  Price
and Swendson use numerical computation as a means for teaching
topics related to ideal quantum gases and describe various projects
that allow students to compute the chemical potential and related
quantities.~\cite{price-swendson}  Mulhall and Moelter calculate 
$N(k)$, use it to compute an approximate density
of states for a rectangular box and a sphere, and then extend the
results to  a system of non-interacting
bosons.~\cite{mulhall-moelter}  Balian and Bloch analyze the
oscillations of the smoothed density of eigenvalues, noting their
important application in nuclear physics.\cite{footnote2}
Cottingham and Greenwood   discuss  the
integrated density of states in a nuclear context, including a plot
comparing the exact (discrete) version to an asymptotic expression
which includes the leading volume and subleading surface
terms.~\cite{cottingham-greenwood}  We also note connections to
the Casimir effect~\cite{casimir} and to the growing body of
literature on nanostructures.~\cite{nanostructures}

The remainder of this paper is organized as follows.  In
Sec.~\ref{sec:distn-eigenvalues} we define the cumulative state
number, $N(k)$, and  give the averaged asymptotic expressions to
which our numerical results will be compared.  We give the
eigenvalues for  the rectangular
parallelepiped, the sphere and the circular cylinder  and show how
the states contributing to $N(k)$ can be visualized as points
on a lattice.  We also derive
the leading and first subleading contributions to the asymptotic expressions for
the rectangular case.  In Sec.~\ref{sec:projects-exercises} we
describe several projects and problems.  Section~\ref{sec:discussion-conclusions} contains some
discussion and concluding remarks.
Appendix~\ref{sec:app-energy-eigenfns} gives expressions for the
wave functions.  In Appendix~\ref{sec:pseudocode} we describe
algorithms that can be used for the various projects, including an 
efficient algorithm  for the spherical case and
pseudocode.  In
Appendix~\ref{sec:derivation-volume-surface} we derive the leading and
first subleading contributions to $\overline{N}(k)$ for the spherical
and cylindrical cases.

\section{Distribution of eigenvalues for three geometrical shapes}
\label{sec:distn-eigenvalues}

Consider Schr\"{o}dinger's equation for a particle that is free
within some bounded domain,
\begin{equation}
  -\frac{\hbar^2}{2\mu} \nabla^2\psi = \epsilon \psi \,,
  \label{eq:TISE}
\end{equation}
where $\epsilon$ is the energy of the particle and $\mu$ is its mass.
Defining the wavenumber as
\begin{equation}
  k = \sqrt{\frac{2\mu \epsilon}{\hbar^2}}
  \label{eq:kdefn}
\end{equation}
immediately gives Eq.~(\ref{eq:wave-equation}).  We
consider a rectangular
parallelepiped with linear dimensions $L_x$, $L_y$ and $L_z$ (in the
$x$, $y$ and $z$ directions, respectively), a sphere of radius $a$ and
a circular cylinder of height $L$ (in the $z$ direction) and radius
$a$.  The particle   is free inside the box in each
case.  At the boundary of the box we assume Dirichlet boundary
conditions; that is, $\psi=0$ at the boundary.  Other boundary conditions
may also be assumed and are appropriate for 
problems such as an electromagnetic wave inside of a cavity, or sound
waves inside an acoustic resonator.~\cite{baltes-hilf}

By solving Eq.~(\ref{eq:wave-equation}) with  the boundary
conditions we find that the  energy eigenvalues
and hence wavenumbers are quantized (see
Appendix~\ref{sec:app-energy-eigenfns}); that is, the wavenumbers form an
(infinite) discrete set. The wavenumbers may be labeled by triplets of integers and thus may
be thought to be on a three-dimensional lattice.  We denote the
eigenvalues  by $k_{\vec{n}}$, where $\vec{n}$ represents
the  triplet of integers labeling the state.  The cumulative
state number, $N(k)$,  counts the number of  distinct states  
for which  $k_{\vec{n}} \leq k$.  That is,\cite{footnote3}
\begin{equation}
  N(k) = \sum_{k_{\vec{n}} \leq k} 1 .
  \label{eq:Nkdefn}
\end{equation}
Weyl showed that, in the limit of large $k$, $N(k)$ is
approximately proportional to the volume, $V$, of the
domain,\cite{weyl}
\begin{equation}
  N(k) \sim \frac{V k^3}{6 \pi^2} + \ldots,
  \label{eq:WeylsThm}
\end{equation}
independent of the  shape of the domain.  This result is
sometimes referred to as Weyl's theorem.  For many applications in
statistical mechanics it is sufficient to keep only the first term in
Eq.~\eqref{eq:WeylsThm}.  Thus, for example, the Fermi energy  is\cite{schroeder}
\begin{equation}
  \epsilon_F  =  \frac{\hbar^2 k^2}{2\mu}  \simeq  
                \frac{\hbar^2}{2\mu}\left(\frac{3\pi^2N_{\rm f}}{V}\right)^{2/3},
\label{eq:fermi-energy}
\end{equation}
where    the number of fermions,
$N_{\rm f}$, is twice $N(k)$, because two spin-1/2
fermions can be in each of the states.
Equation~(\ref{eq:fermi-energy})  holds only for large $k$; for
smaller $k$ we must be 
careful (see Ref.~\onlinecite{price-swendson} for a discussion of the case $N_{\rm f}=1$).  The density of
states, which is the number of single-particle states per unit energy,
is\cite{footnote4}
\begin{equation}
  g(\epsilon) = \frac{dN}{dk} \frac{dk}{d\epsilon} \simeq
    \frac{V \mu^{3/2}}{\sqrt{2}\pi^2\hbar^3}\epsilon^{1/2} \; ,
\end{equation}
where we have used Eq.~(\ref{eq:kdefn}).

Many corrections to Eq.~(\ref{eq:WeylsThm}) have been computed over
the years.  As noted in Ref.~\onlinecite{baltes-hilf}, the difference
between $N(k)$ and its asymptotic approximations exhibits fluctuations
that   disguise the geometric nature of the approximation.  To gain  insight into the structure of the cumulative
state number, it is necessary to perform some   averaging over
the eigenvalues.  This averaging can also be interpreted physically because, for example, we do not have measuring
devices that can measure with perfect accuracy, and the walls of
bounding domains are not perfectly rigid or smooth.~\cite{baltes-hilf}  In
practice, these effects lead to a natural sort of averaging.  There is
an extensive literature on how    to average over the
eigenvalues.~\cite{baltes-hilf}  Having performed the averaging in an
appropriate manner, we can make rigorous statements about the
behavior of  $\overline{N}(k)$   as a function of $k$.  The averaged asymptotic
expression for $\overline{N}(k)$ for
the cases  we consider is given by~\cite{baltes-hilf, gutierrez-yanez,
  balian-bloch, waechter}
\begin{equation}
  \overline{N}(k) \sim \frac{V k^3}{6 \pi^2} -\frac{Sk^2}{16\pi} +\frac{Mk}{6\pi^2}+\frac{J}{512\pi}+\ldots
  \label{eq:Navge-four-terms}
\end{equation}
for Dirichlet boundary conditions; $S$ is the surface area of
the bounding domain.  For smooth boundaries, the quantities $M$ and
$J$ may be expressed as surface integrals that depend on the radii of
curvature of the domain.\cite{footnote5} Expressions for the
rectangular parallelepiped, sphere and cylinder are  found in
Table~\ref{tab:MJ}.

\begin{table}
\centering
    \begin{tabular}{lcc}
      \hline\hline
      & $M$ & $J$ \\
      \hline
      Rectangular Parallelepiped~~~~~~ & $\dfrac{3\pi}{2}\left(L_x +L_y + L_z\right)$ & $-64\pi$ \\
      Sphere & $4\pi a$ & $0$ \\
      Circular Cylinder & $\dfrac{\pi}{2}\left(2L+3\pi a\right)$ & $2\pi\left(\dfrac{L}{a} -\dfrac{64}{3}\right)$ \\
      \hline\hline
    \end{tabular}
    \caption{The constants $M$ and $J$ in
      Eq.~(\ref{eq:Navge-four-terms}) for the rectangular
      parallelepiped, sphere and circular cylinder.~\cite{baltes-hilf,
        gutierrez-yanez, waechter}}
    \label{tab:MJ}
\end{table}

In the following  we give  expressions for the
eigenvalues for the three geometrical domains   and     the cumulative state number $N(k)$.
We also give a quasi-rigorous derivation of the first two terms in
Eq.~(\ref{eq:Navge-four-terms}) for  the rectangular
parallelepiped.  The  derivations for the spherical and
cylindrical cases  are given in
Appendix~\ref{sec:derivation-volume-surface}.  In principle we can
also calculate the density of states numerically.  The
density of states is a sum of delta functions corresponding to the
slopes of the steps in $N(k)$ (see  Ref.~\onlinecite{balian-bloch}),
but we can smooth the distribution in various
ways.\cite{baltes-hilf, mulhall-moelter, balian-bloch}  Because the
density of states involves a derivative, it is a
noisier quantity,\cite{mulhall-moelter}  and thus we will consider only the cumulative state number.

\subsection{Rectangular parallelepiped}
\label{sec:rectangular-parallelepiped}

The eigenvalues for the rectangular parallelepiped may be
written as
\begin{equation}
  k^2_{n_xn_yn_z} = \pi^2\left(\frac{n_x^2}{L_x^2}
  +\frac{n_y^2}{L_y^2}+\frac{n_z^2}{L_z^2}\right) \, 
  \label{eq:kvecndefn}
\end{equation}
(see Appendix~\ref{sec:app-rect-energy-eigenfns}), so that $N(k)$ sums
up the states satisfying
\begin{equation}
  \pi^2\left(\frac{n_x^2}{L_x^2}
  +\frac{n_y^2}{L_y^2}+\frac{n_z^2}{L_z^2}\right)
   \leq k^2 \, ,
  \label{eq:nxnynzbdy}
\end{equation}
where $n_x$, $n_y$ and $n_z$  each independently take on the values
1, 2, 3,\ldots.

%
\begin{figure}[t]
\centering
\resizebox{4in}{!}{\includegraphics*{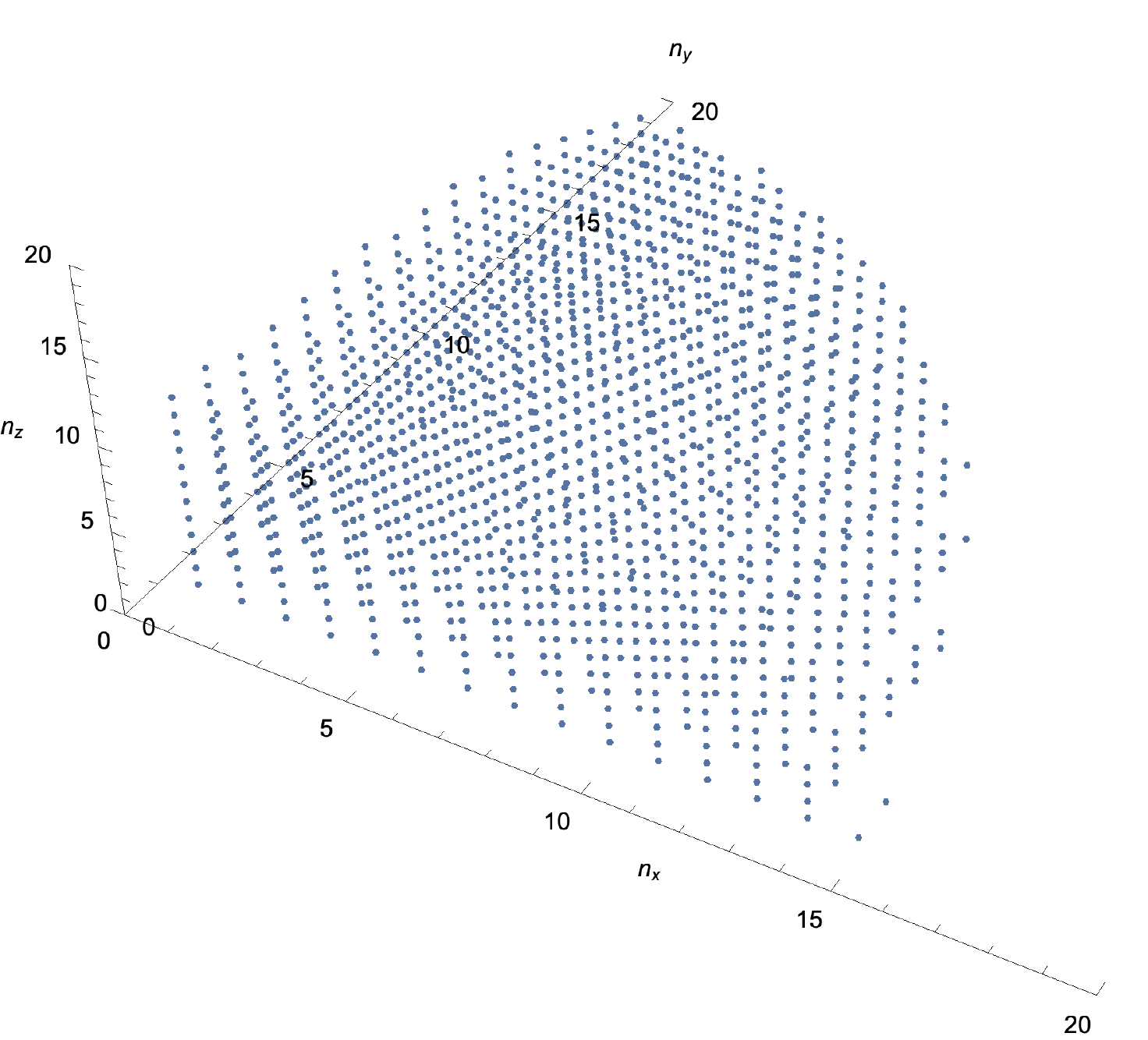}}
\caption{States contributing to $N(k)$ for $kV^{1/3}=15\pi$ for a
  rectangular parallelepiped with $L_x=1.25L_z$ and $L_y=1.5L_z$.
  $N(k)$ is  the total number of  points in the plot which,
  in this case, is 1505.  Only positive integers $n_x$,
  $n_y$ and $n_z$ are included in the sum.}
\label{fig:nxnynz}
\end{figure}

Figure~\ref{fig:nxnynz} shows the states contributing to $N(k)$ for
  $kV^{1/3}=15\pi$, with $L_x=1.25L_z$ and $L_y=1.5L_z$.  The
inequality in Eq.~(\ref{eq:nxnynzbdy}) shows that the bounding surface
is an ellipsoid in $n_x n_y n_z$ space.  Let us imagine attempting
to count the points.  Because the integers $n_x$, $n_y$ and $n_z$ are
all    positive, we first count all of the points inside
the entire bounding ellipsoid (including points with $n_x$, $n_y$ or $n_z$ equal to zero),
and then subtract  the points in the $n_x n_y$, $n_yn_z$
and $n_xn_z$ planes.  Having done so, we will have
subtracted  the points along each of the axes twice, so they 
need to be added back in once.  Finally, the origin  needs to be
subtracted  out once, and the result divided by 8.  The number of
points can thus be written as\cite{footnote6}
\begin{equation}
  N_{\rm rect}(k) = \frac{1}{8}\left[A_3(k) - A_2^{xy}(k)-A_2^{yz}(k)-A_2^{xz}(k)+A_1^x(k)+A_1^y(k)+A_1^z(k)-1\right] \,,
  \label{eq:NkA}
\end{equation}
where $A_3$ is the sum of all points inside the ellipsoidal volume, 
$A_2^{ij}$ represents the sums bounded by ellipses in the various planes,
and  $A_1^i$ represents the sums along the axes.

For $k \gg 1/L_i$, the sum represented by $A_3(k)$
may be approximated by an integral.  Because the points in
$\vec{n}$-space are uniformly spaced (with unit density), $A_3(k)$ is
approximately equal to the volume of the bounding ellipsoid,
\begin{equation}
\dfrac{4}{3}\pi
\left(\dfrac{kL_x}{\pi}\right)\left(\dfrac{kL_y}{\pi}\right)\left(\dfrac{kL_z}{\pi}\right)
= 4k^3L_xL_yL_z/(3\pi^2),
\end{equation}
or
\begin{equation}
  A_3(k) = \frac{4Vk^3}{3\pi^2}+P_3(k),
  \label{eq:A3}
\end{equation}
where $V=L_xL_yL_z$ 
and $P_3(k)$ is the error incurred by
approximating the sum of lattice points by an integral.~\cite{baltes-hilf}
We can proceed   similarly for  $A_2^{ij}$ by approximating the sums
in the various planes by the areas of the corresponding ellipses to find
\begin{equation}
  A_2^{ij}(k) = \frac{1}{\pi}L_iL_jk^2 +P_2^{ij}(k).
  \label{eq:A2ij}
\end{equation}
If we substitute Eqs.~(\ref{eq:A3}) and (\ref{eq:A2ij}) into
Eq.~(\ref{eq:NkA}), drop the remainders and note that
$S=2(L_xL_y+L_yL_z+L_xL_z)$,  we can reproduce the first
two terms in Eq.~(\ref{eq:Navge-four-terms}).

Our derivation of the
surface area term in Eq.~(\ref{eq:Navge-four-terms}) has  been
 heuristic.  In particular, we have implicitly assumed that
the lattice remainder from the volume term is smaller than $O(k^2)$,
so that it does not overwhelm our calculation of the surface area
term.  Reference~\onlinecite{baltes-hilf}   discusses lattice remainders and the asymptotic expression for $N$;
the result that we have derived is
accurate when the averaging is done  appropriately.


Suppose that instead of doing the volume
integral over the entire ellipsoid and then dividing by eight, we had
 done the volume integral over one eighth of the ellipsoid.
This integral would have included contributions from the points on the
lattice on each of the planes $n_x=0$, $n_y=0$ and $n_z=0$.  The area
of each of these contributions is one quarter of the area of the
corresponding ellipse represented by $A_2^{ij}$.  And yet we see that
we must subtract  one eighth this area, not one fourth.
Evidently the one eighth volume integral, with its bounding
surfaces precisely at $n_x=0$, etc., effectively captures a number of
lattice points equal to one half the surface area of each of these
bounding surfaces; this is  the amount that must be subtracted.  Alternative approaches are  to perform the volume integration
in one eighth of the space, but starting from $n_x, n_y, n_z =1$ (in
which case an area integral must be  added to the result to
correct for the fact that only half the contributions from the
bounding surfaces at $n_x, n_y, n_z =1$ are  included in the
volume integral) or  to perform the volume integral starting from
$n_x, n_y , n_z =1/2$ (in which case no area correction needs to
be done).  All of these approaches yield the same result to order
$k^2$.

To compare different domain shapes on the same
plot, it is convenient to define the following dimensionless
variables,
\begin{align}
  \kappa & =  kV^{1/3} \label{eq:kappadefn} \\
  \xi_x & =  L_x/L_z \label{eq:xix} \\
  \xi_y & =  L_y/L_z \label{eq:xiy} \,,
\end{align}
so that
\begin{equation}
  \kappa^2_{n_xn_yn_z} = \pi^2\left[n_x^2\left(\frac{\xi_y}{\xi_x^2}\right)^{2/3}+n_y^2\left(\frac{\xi_x}{\xi_y^2}\right)^{2/3}+
  n_z^2\left(\xi_x\xi_y\right)^{2/3}\right] \; .
  \label{eq:kappanxnynz}
\end{equation}
We can  now  write Eq.~(\ref{eq:Navge-four-terms}) for the
rectangular parallelepiped as
\begin{equation}
  \overline{N}_{\rm rect}(\kappa)
   \sim  \frac{\kappa^3}{6 \pi^2} - \kappa^2f_{\rm rect}^{(2)}(\xi_x,\xi_y)
     + \kappa f_{\rm rect}^{(1)}(\xi_x,\xi_y)  -\frac{1}{8}+\ldots \, ,
  \label{eq:N-rect-four-terms}
\end{equation}
where
\begin{align}
  f_{\rm rect}^{(1)}(\xi_x,\xi_y)
  & =  \frac{1}{4\pi}\left[\left(\frac{\xi_x^2}{\xi_y}\right)^{1/3}+
        \left(\frac{\xi_y^2}{\xi_x}\right)^{1/3}+\left(\frac{1}{\xi_x\xi_y}\right)^{1/3}\right] \label{eq:f1rect} \\
  f_{\rm rect}^{(2)}(\xi_x,\xi_y)
  & =  \frac{1}{8\pi}\left[\left(\xi_x\xi_y\right)^{1/3}+
      \left(\frac{\xi_x}{\xi_y^2}\right)^{1/3}+\left(\frac{\xi_y}{\xi_x^2}\right)^{1/3} \right]\,. \label{eq:f2rect}
\end{align}
Equation~(\ref{eq:N-rect-four-terms}) gives the averaged asymptotic
expression to which we can compare our exact (discrete) results.
Note that the leading term is  independent of the shape of
the rectangular box and has the same form for a cube as it does for
a quasi-one- or two-dimensional box.  The shape dependence only shows
up in the subleading terms, which depend on $\xi_x$ and $\xi_y$.

\begin{figure}[t]
\centering
\resizebox{6in}{!}{\includegraphics*{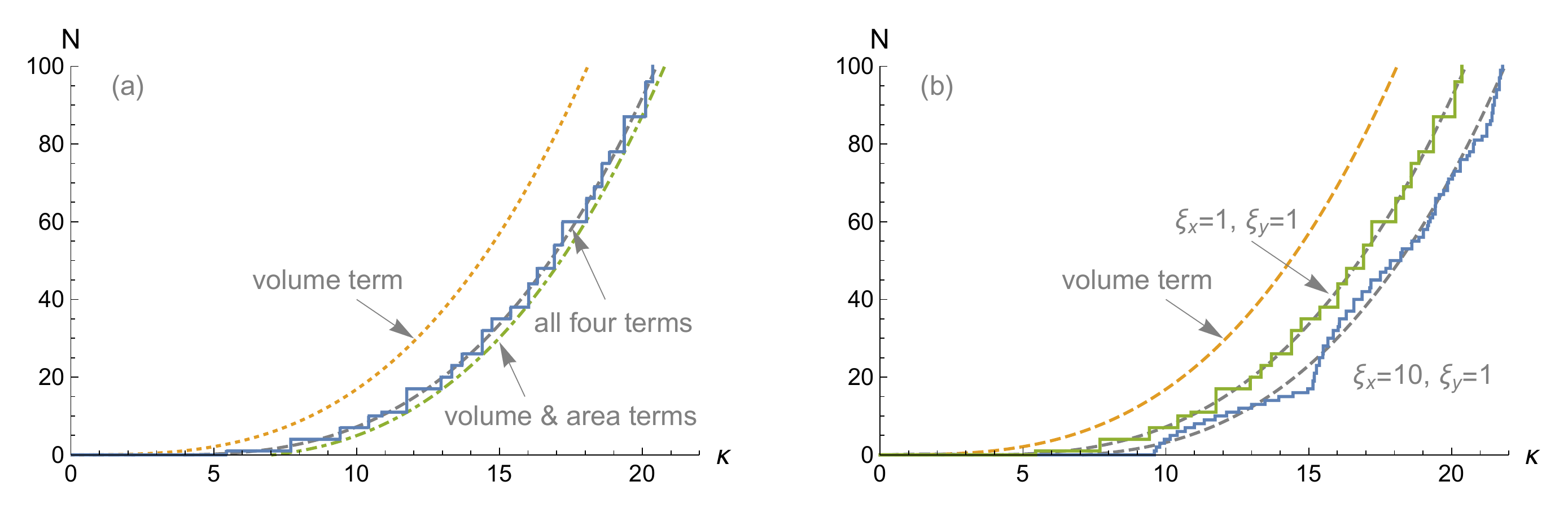}}
\caption{$N$ versus $\kappa=kV^{1/3}$ for two 
  rectangular domains.   (a)  $(\xi_x,\xi_y) = (1,1)$,
  where $\xi_{x,y}=L_{x,y}/L_z$.  The (blue) solid line shows the 
  numerical data, which increases in discrete steps as new states are
  added.  Also shown are various  approximations from
  Eq.~(\ref{eq:N-rect-four-terms}): the volume term  (proportional
  to $\kappa^3$), the volume and area terms, and all four terms.  
  (b)  Comparison of $(\xi_x,\xi_y)=(1,1)$ (a
  cube) and $(\xi_x,\xi_y)=(10,1)$ (a quasi-one-dimensional box).  The
  dashed lines show the theoretical approximation from
  Eq.~(\ref{eq:N-rect-four-terms}).}
\label{fig:Nparallelepiped}
\end{figure}

Figure~\ref{fig:Nparallelepiped} shows $N(\kappa)$ as a function of
$\kappa$ for two  rectangular parallelepipeds.  The exact
results consist of a series of discrete steps corresponding to the
addition of new states at various values of $\kappa$.  We have derived
$N(\kappa)$  following the algorithm outlined in
Appendix~\ref{sec:pseudocode-rect}, which essentially corresponds to
adding up the number of states in a region similar to that shown in
Fig.~\ref{fig:nxnynz}.\cite{footnote7}  Most textbook derivations of the
cumulative state number   include only the leading volume
term.  As is evident from Fig.~\ref{fig:Nparallelepiped}, the
volume term in Eq.~(\ref{eq:N-rect-four-terms}) (proportional to
$\kappa^3$) does not give a  good estimate for $N(\kappa)$
for small $\kappa$.  Inclusion of the negative surface area
term (proportional to $\kappa^2$) improves the agreement
significantly, but overshoots somewhat.  Inclusion of all four terms
in Eq.~(\ref{eq:N-rect-four-terms}) leads to  good agreement on
average, although the exact result  fluctuates about the
asymptotic curve.  The corrections
to the leading term are of  order $\kappa^2$ and become
unimportant as $\kappa\to \infty$.  We note that the approach to the
asymptotic behavior is much slower for the quasi-one-dimensional
structure in Fig.~\ref{fig:Nparallelepiped}(b) (that is, the curve is
displaced further to the right) than for the cubic structure.

The quasi-one-dimensional case in Fig.~\ref{fig:Nparallelepiped}(b)
also exhibits cusps.  To understand these, note that the $n_x^2$
coefficient in Eq.~(\ref{eq:kappanxnynz}) is much smaller than the
$n_y^2$ and $n_z^2$ coefficients in this case.  Thus, between
$\kappa\approx 9$ and $\kappa\approx 15$, $n_y=n_z=1$ and $N=n_x$.
Indeed, $N(\kappa)$ between $\kappa\approx 9$ and $\kappa\approx 15$
is relatively well-described by a continuous curve that depends only
on $n_x$.  For $\kappa \gtrsim 15$, states with $(n_y,n_z)=(1,2)$ and
$(2,1)$ also start to contribute to $N(\kappa)$, leading to an abrupt
change in slope of $N(\kappa)$.  The quasi-two-dimensional case also
exhibits cusps, and the behavior of $N(\kappa)$ at and between the
cusps may similarly be understood by considering quantum numbers
associated with the large and small directions to behave
quasi-continously and discretely, respectively.  The situation here is
reminiscent of the Casimir effect, in which a sum over energy states
can be written in terms of a mixture of continuous and discrete
variables.~\cite{elizalde-romeo}  Further analysis of quasi-one- and
two-dimensional systems is left for Problems~1.2, 1.3, 3.2 and 3.3.

\begin{figure}[t]
\centering
\resizebox{6in}{!}{\includegraphics*{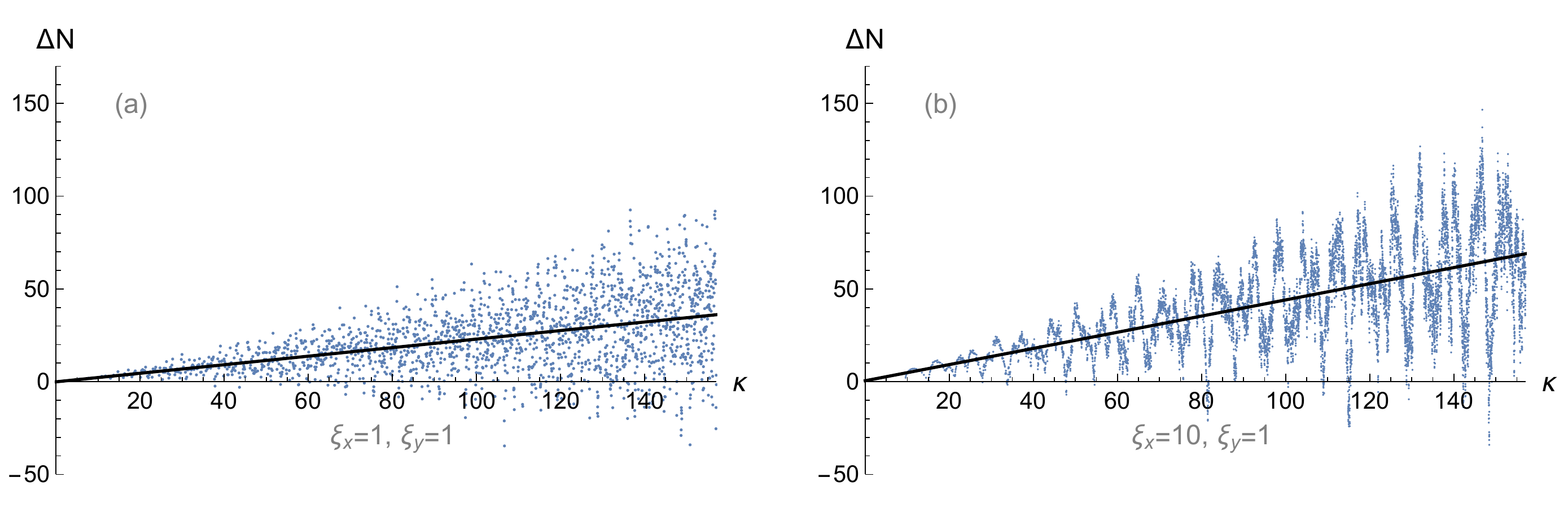}}
\caption{$\Delta N_{\rm rect}$  defined in Eq.~(\ref{eq:DeltaN-rect}) versus $\kappa$
  for rectangular domains with (a) $(\xi_x,\xi_y) = (1,1)$ and (b)
  $(\xi_x,\xi_y) = (10,1)$.  The slopes of the best-fit
  lines give the coefficients of the linear terms in
  Eq.~(\ref{eq:N-rect-four-terms}).  The plot in (b) has far
  more points than that in (a) because the case
  $(\xi_x,\xi_y) = (1,1)$ has many degeneracies.}
\label{fig:DeltaNparallelepiped}
\end{figure}
%

To see the fluctuations about the asymptotic expression more
clearly, we  subtract the asymptotic expression from our exact
(discrete) result for $N(k)$.  If we
define
\begin{equation}
  \Delta N_{\rm rect} (\kappa) \equiv N_{\rm rect}(\kappa) 
  -\frac{\kappa^3}{6 \pi^2} + \kappa^2f_{\rm rect}^{(2)}(\xi_x,\xi_y) \, ,
  \label{eq:DeltaN-rect}
\end{equation}
the result should be approximately linear in $\kappa$,
\begin{equation}
  \Delta N_{\rm rect} (\kappa) \sim
  \kappa f_{\rm rect}^{(1)}(\xi_x,\xi_y)  -\frac{1}{8} \, ,
\end{equation}
and we can perform a fit to determine the linear coefficient,
$f_{\rm rect}^{(1)}(\xi_x,\xi_y)$ using our numerical
data.  Figure~\ref{fig:DeltaNparallelepiped} shows plots of
$\Delta N_{\rm rect}$; the results
exhibit large fluctuations, as expected.  Although the spread of
$\Delta N_{\rm rect}$ values looks  large on
the scale used in Fig.~\ref{fig:DeltaNparallelepiped}, note that the vertical scale for
$N_{\rm rect}$ would have a maximum of order $6\times10^4$
for the same range of $\kappa$ values.

The asymptotic
expression in Eq.~(\ref{eq:N-rect-four-terms}) implicitly assumes that
an averaging procedure has been performed (see
Ref.~\onlinecite{balian-bloch}, for example).  Rather than implement an
 involved averaging procedure, we adopt a 
simple-minded approach and simply fit a line to the points in the
plots.  Table~\ref{tab:DeltaN} compares the slopes in
Fig.~\ref{fig:DeltaNparallelepiped} to the theoretical
values.  The agreement is  good.  To obtain good agreement, we  had to implement a slightly
improved prescription for $N$,\cite{baltes-hilf}
\begin{equation}
  N(k) = \sum_{k_{\vec{n}} < k} 1 + \sum_{k_{\vec{n}} = k} \frac{1}{2} \,
  \label{eq:Nkdefn-improved}
\end{equation}
[compare with Eq.~(\ref{eq:Nkdefn})].  This revised prescription is
particularly important for   $(\xi_x,\xi_y) = (1,1)$, which has
large degeneracies.  Using the original prescription in this case
yields a linear coefficient that is more than 50\% higher than the
theoretical one.

\begin{table}[t]
\centering
    \begin{tabular}{lcc}
      \hline\hline
      & $f_{\rm rect}^{(1)}(\xi_x,\xi_y)$ & Linear Fit \\
      \hline
      $(\xi_x,\xi_y) = (1,1)$~~~~~~ & ~~~~~~~0.239~~~~~~~ & ~~~~~~~0.230~~~~~~~ \\
       $(\xi_x,\xi_y) = (10,1)$ & 0.443 & 0.436 \\
      \hline\hline
    \end{tabular}
    \caption{Comparison of  theoretical and fitted values for the
      coefficient of the linear term in
      Eq.~(\ref{eq:N-rect-four-terms}) for the rectangular
      parallelepiped.  The     second column gives
      the slopes of the solid lines in
      Fig.~\ref{fig:DeltaNparallelepiped}.  The fits include small
      constant terms as well, but these constants do not agree 
      well with the expected value of $-1/8$.}
    \label{tab:DeltaN}
\end{table}

\subsection{Sphere}
\label{sec:spherical}

For a sphere of radius $a$ with Dirichlet boundary conditions, the  
wavenumbers are given by
\begin{equation}
  k_{n \ell m}^2 = \frac{\beta_{\ell,n}^2}{a^2},
  \label{eq:knlm}
\end{equation}
where $\beta_{\ell,n}$ denotes the $n$th zero of the spherical Bessel
function, $j_\ell$; that is, $j_\ell(\beta_{\ell,n})=0$ (see
Appendix~\ref{sec:app-sphere-energy-eigenfns}).  The integer $\ell$ is
related to the orbital angular momentum of the state and $m$ is
related to its azimuthal component.  The quantum numbers take on the
values $n=1, 2, 3,\ldots$, $\ell=0,1,2,3,\ldots$ and
$m=-\ell, -\ell+1, \ldots, \ell-1, \ell$.  The cumulative state number $N(k)$  counts  all of the states for which
\begin{equation}
  \frac{\beta_{\ell,n}^2}{a^2} \leq k^2 \;.
  \label{eq:knlmineq}
\end{equation}
As is evident from Eq.~(\ref{eq:knlm}), the quantized wavenumbers have
a $(2\ell+1)$-fold degeneracy, because the right-hand side  does not depend on $m$.  Thus, in actual computations of
$N(k)$, we need only loop over $\ell$ and $n$ (being careful to keep
track of the degeneracy).

\begin{figure}[t]
\centering
  \resizebox{4in}{!}{\includegraphics*{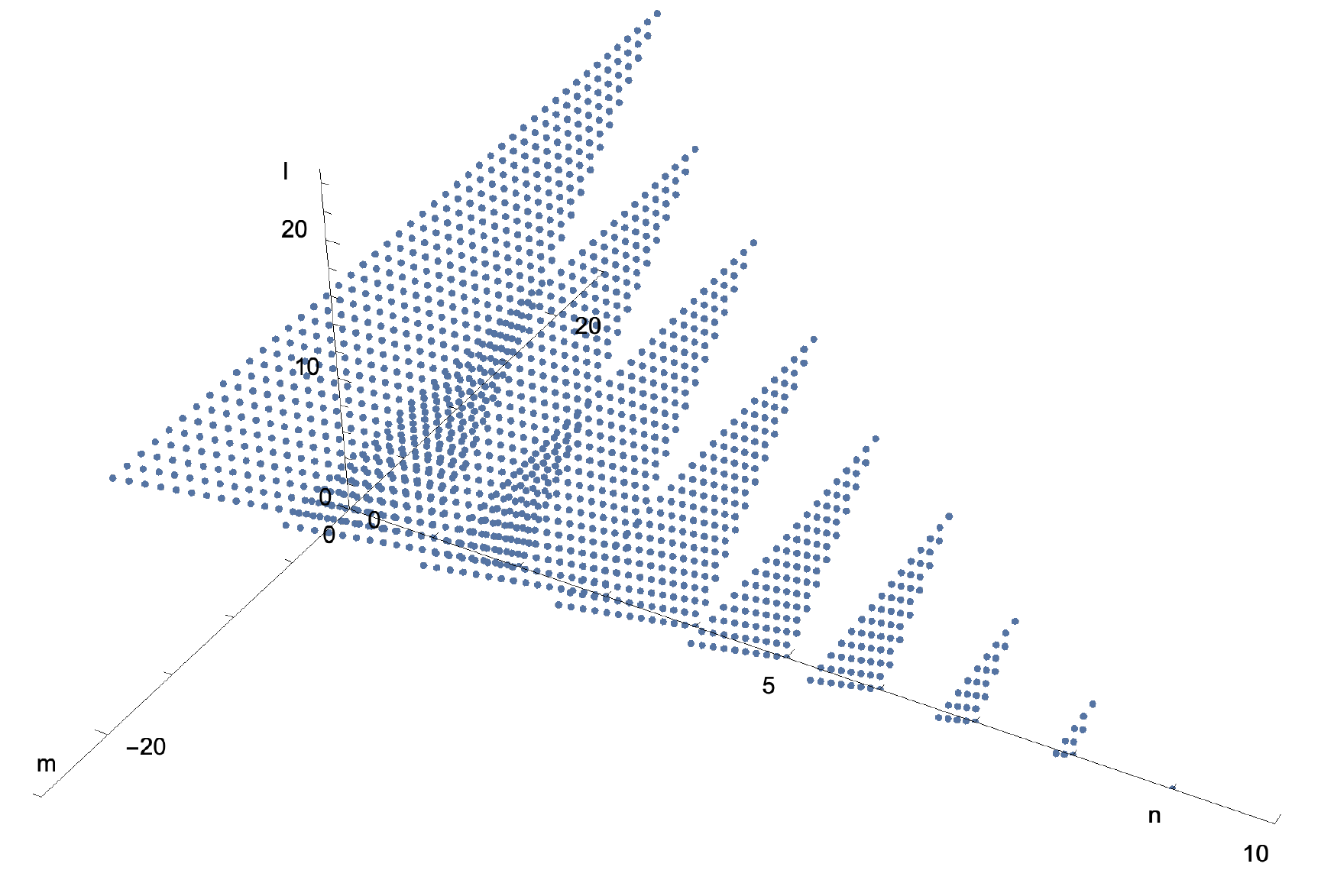}}
  \caption{States contributing to $N(k)$ for a spherical volume with
    $kV^{1/3}=15\pi$.  $N(k)$ is  the sum of the 1561 points shown in
    the plot.   The scales are
    different for the different directions, which gives the impression
    that the points are not uniformly spaced (which they
    are).}
\label{fig:nml}
\end{figure}

It is useful to visualize
the quantum numbers $n$, $\ell$ and $m$ in terms of a three-dimensional
lattice, as we did for the rectangular case.  This approach is helpful
for determining the volume and area terms in
Eq.~(\ref{eq:Navge-four-terms}) (see
Appendix~\ref{sec:derivation-volume-surface-sphere}) and for
visualizing the algorithm used to compute the cumulative state number.
Figure~\ref{fig:nml} shows the states contributing to $N(k)$ for 
 $kV^{1/3}=15\pi$ (which is the same value that was used for
Fig.~\ref{fig:nxnynz}).

%

\begin{figure}[t]
\centering
  \resizebox{4in}{!}{\includegraphics*{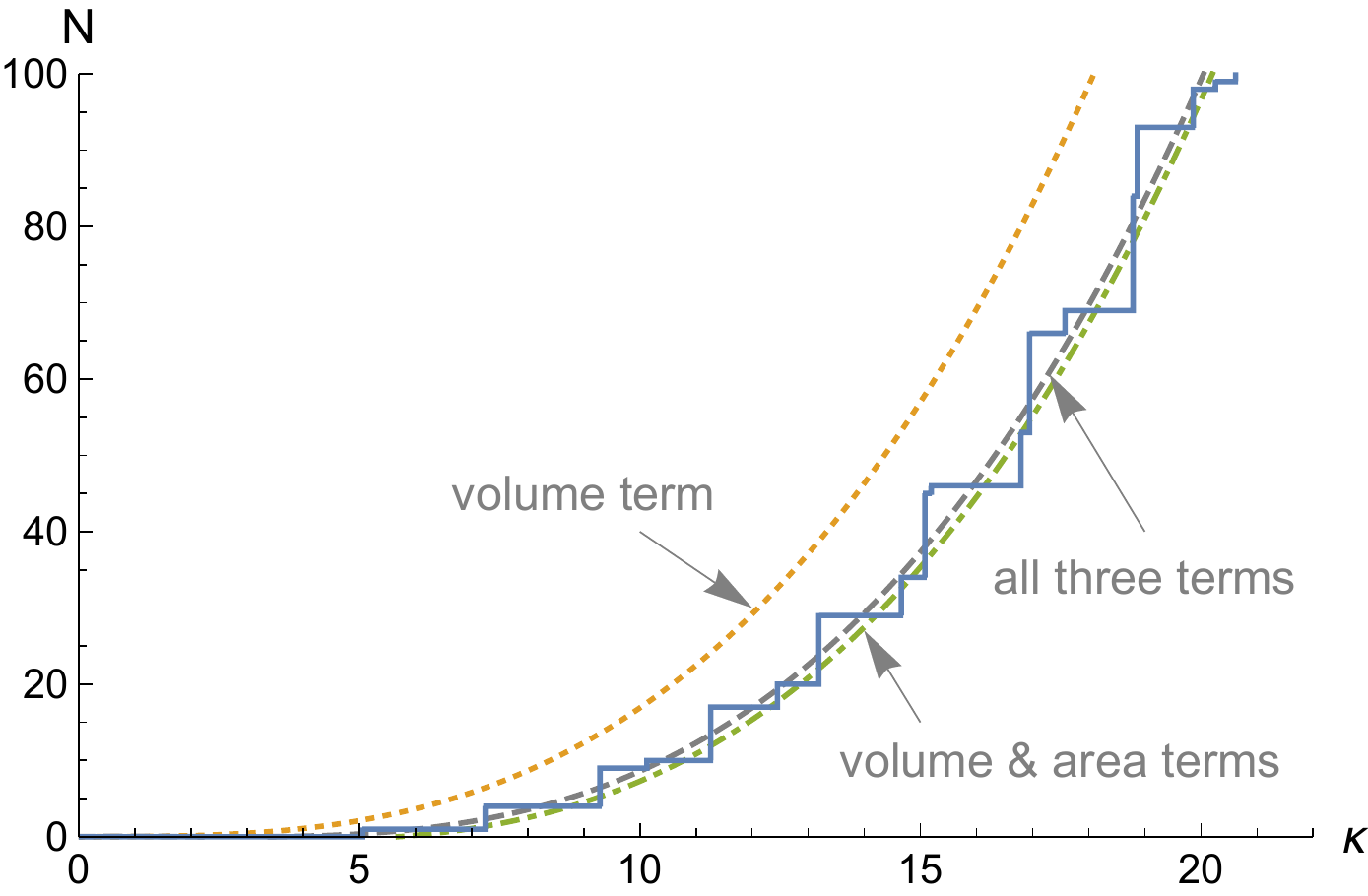}}
  \caption{$N(\kappa)$ for a sphere.  The smooth dashed lines
    show the theoretical result from
    Eq.~(\ref{eq:N-sphere-four-terms}), including one, two and three
    terms.  Comparison with the analogous plots in
    Fig.~\ref{fig:Nparallelepiped} shows that the discrete steps
    are typically much larger in the present case due to the
    high degree of symmetry and the resulting
    $(2\ell+1)$-fold degeneracies, which become large for large $\ell$.
    This effect also leads to large fluctuations when we calculate
    $\Delta N_{\rm sph}$.}
\label{fig:Nk-sphere}
\end{figure}

We define the dimensionless variable $\kappa$ as in Eq.~(\ref{eq:kappadefn}) and express
the eigenvalues   as
\begin{equation}
  \kappa_{n\ell m}^2 = \left(\frac{4\pi}{3}\right)^{2/3}\beta_{\ell,n}^2 \,.
  \label{eq:kappanlm}
\end{equation}
The asymptotic expression for the averaged cumulative state number
becomes (see Table~\ref{tab:MJ})
\begin{equation}
  \overline{N}_{\rm sph}(\kappa)
   \sim  \frac{\kappa^3}{6 \pi^2} - \left(\frac{3}{32\pi}\right)^{2/3}\kappa^2
           + \left(\frac{2}{9\pi^4}\right)^{1/3}\kappa + 0 +\ldots \,.
  \label{eq:N-sphere-four-terms}
\end{equation}

Figure~\ref{fig:Nk-sphere} shows a plot of $N(\kappa)$ as a function
of $\kappa$ for a sphere, with smooth curves showing the inclusion of
various terms from Eq.~(\ref{eq:N-sphere-four-terms}).  The plot
looks  similar to Fig.~5.2 in Ref.~\onlinecite{cottingham-greenwood},
which was considered in the context of a discussion on nuclear models.
We can also define $\Delta N_{\rm sph}$ in analogy
with Eq.~(\ref{eq:DeltaN-rect}) and use the result to determine the
linear coefficient.  This analysis is the subject of Projects~2.2 and
2.3.  The fluctuations are very large in this case, due to the high
degree of symmetry and the resulting large degeneracies in the
eigenvalues.\cite{baltes-hilf}

\subsection{Circular cylinder}
\label{sec:cylindrical}

\begin{figure}[t]
\centering
  \resizebox{4in}{!}{\includegraphics*{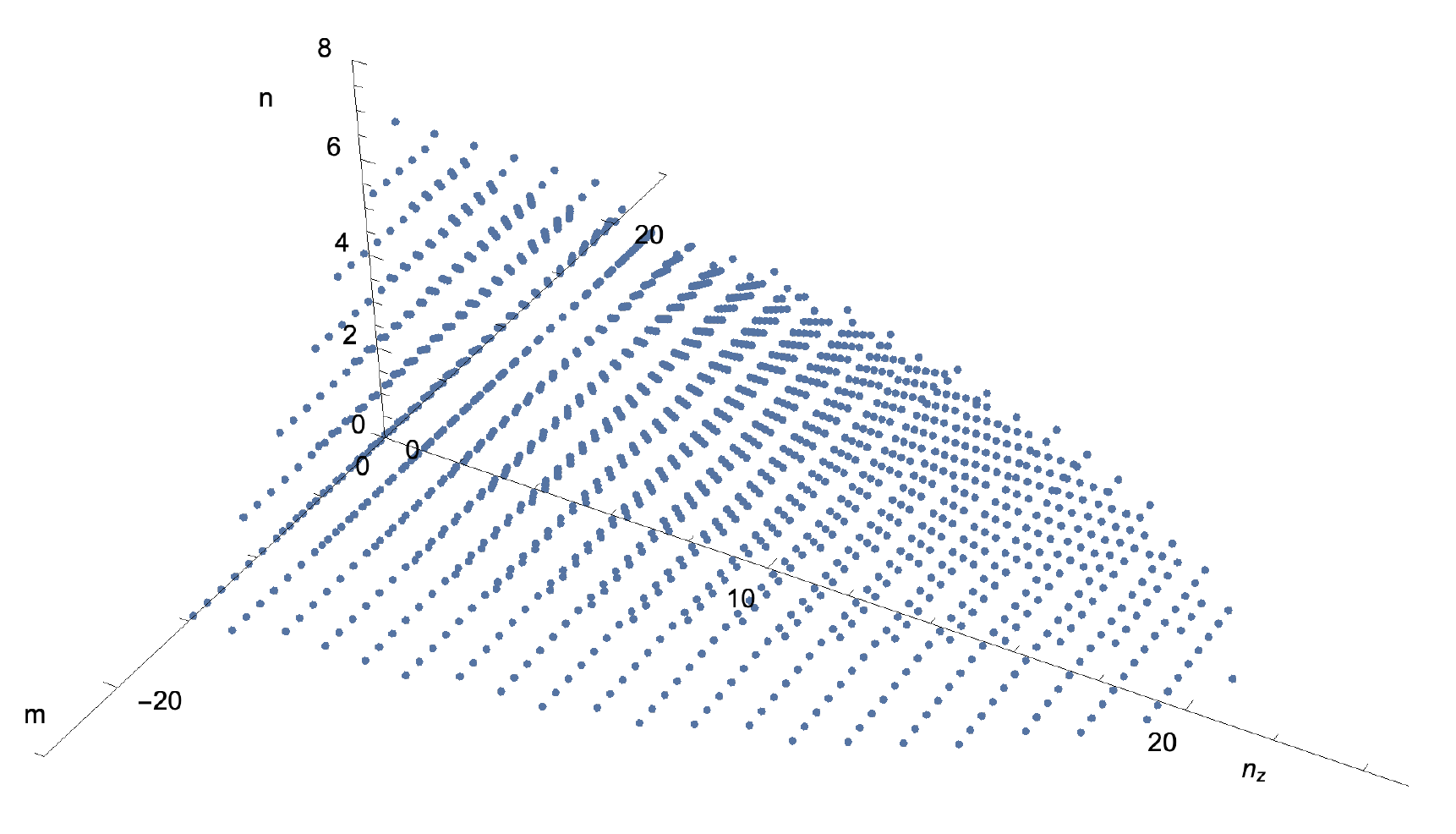}}
  \caption{The 1529 states contributing to $N(k)$ for a cylindrical volume
    with $kV^{1/3}=15\pi$ and $a=L/3$.   Note that $n$ and $n_z$ both start at one,
    while $m$ includes   positive and negative integers,
    as well as zero.}
\label{fig:nzmn}
\end{figure}

The  wavenumbers for a circular cylinder of radius $a$ and
length $L$ with Dirichlet boundary conditions are given by
\begin{equation}
  k_{nmn_z}^2 = \left(\frac{\zeta_{m,n}}{a}\right)^2 + \left(\frac{\pi n_z}{L}\right)^2  \; ,
  \label{eq:knmnz}
\end{equation}
where $\zeta_{m,n}$ denotes the $n$th zero of the regular Bessel
function $J_m$; that is, $J_m(\zeta_{m,n})=0$ (see
Appendix~\ref{sec:app-cylinder-energy-eigenfns}).  Also,
$n=1,2,3,\ldots$, $n_z = 1, 2, 3,\ldots$, and
$m=0, \pm1, \pm2, \ldots$.  Figure~\ref{fig:nzmn} illustrates the
states contributing to $N(k)$ for $kV^{1/3}=15\pi$, with $a=L/3$.  The
points corresponding to these states are uniformly spaced in
$n_z$-$m$-$n$ space and represent all of the states for which
$k_{nmn_z}^2\leq (15\pi)^2/V^{2/3}$.  If we compare the values of
 $N(k)$ for the parallelepiped, sphere and circular cylinder for
the same value of $kV^{1/3}$ (see Figs.~\ref{fig:nxnynz},
\ref{fig:nml} and \ref{fig:nzmn}), we see that  $N(k) \approx 1500$
in each case, in accordance with Weyl's theorem.

%

We define $\kappa$ as in
Eq.~(\ref{eq:kappadefn}) and express the eigenvalues as
\begin{equation}
  \kappa_{nmn_z}^2 = \left(\frac{\pi}{\xi_a}\right)^{2/3}\zeta_{m,n}^2+\left(\pi^4\xi_a^2\right)^{2/3}n_z^2 \,,
  \label{eq:kappanmnz}
\end{equation}
where we have defined $\xi_a$ as
\begin{equation}
  \xi_a = \frac{a}{L} \,.
  \label{eq:xia}
\end{equation}
The asymptotic expression for the averaged cumulative state
number becomes
\begin{equation}
  \overline{N}_{\rm cyl}(\kappa)
   \sim  \frac{\kappa^3}{6 \pi^2} - \kappa^2f_{\rm cyl}^{(2)}(\xi_a)
           + \kappa f_{\rm cyl}^{(1)}(\xi_a)
           +\frac{1}{256}\left(\frac{1}{\xi_a}-\frac{64}{3}\right)+\ldots \, ,
  \label{eq:N-cyl-four-terms}
\end{equation}
where
\begin{align}
  f_{\rm cyl}^{(1)}(\xi_a)
  & =  \frac{1}{4\pi}\left[\frac{2}{3}\left(\frac{1}{\pi\xi_a^2}\right)^{1/3}+
        \left(\pi^2\xi_a\right)^{1/3} \right]\label{eq:f1cyl} \\
  f_{\rm cyl}^{(2)}(\xi_a)
  & =  \frac{1}{8\pi}\left[\left(\pi\xi_a^2\right)^{1/3}+
      \left(\frac{\pi}{\xi_a}\right)^{1/3} \right] \,. \label{eq:f2cyl}
\end{align}

\section{Projects and problems}
\label{sec:projects-exercises}

In this section we outline several projects and problems.  The projects
are primarily computational in nature, and the problems focus
on derivations and  theoretical analyses of numerical results.  The
projects can  be done without the problems, but the reverse is not
true.

The projects and problems  are more or less
self-contained, although we recommend that  Project~1.1
be done first.  The code from this project can then be modified for the
spherical and cylindrical cases, which have the  complication
that students need to determine the zeros of Bessel functions.  For
each geometry there is a project that focuses on determining the
linear coefficient in the asymptotic expression for
$\overline{N}(\kappa)$.  This determination is  challenging
for the sphere, because the amplitude of the fluctuations is
 large in this case.  Measurement of the linear coefficient for
the spherical case spans Projects~2.2 and 2.3. In the first,
students implement a much more efficient algorithm for determining
$N(\kappa)$.  In the second, they average $\Delta N$ over small
intervals and perform a weighted linear fit to the results.

In the following  ``calculate
$N(\kappa)$ as a function of $\kappa$''    means to
determine a discrete set of ordered pairs $(\kappa,N(\kappa))$ for the
eigenvalues $\kappa$.  This determination is not to be confused with calculating the
corresponding averaged, asymptotic expression, $\overline{N}(\kappa)$,
which is a continuous function of $\kappa$.

\subsection{Rectangular parallelepiped}

\noindent {\bf Problem 1.1 -- Theoretical expressions}

(a) Start from Eq.~(\ref{eq:kvecndefn}) and
use the definitions of the dimensionless variables in
Eqs.~(\ref{eq:kappadefn})--(\ref{eq:xiy}) to derive
Eqs.~(\ref{eq:kappanxnynz}) and (\ref{eq:N-rect-four-terms}).

(b) Determine the values of $\xi_x$ and $\xi_y$ that minimize
$f_{\rm rect}^{(1)}(\xi_x, \xi_y)$ and
$f_{\rm rect}^{(2)}(\xi_x, \xi_y)$.  What shape is
the bounding domain in this case?

\bigskip
\noindent
{\bf Project 1.1 -- $\bm{N(\kappa)}$ for a cube}

\noindent
Use the approach outlined in Appendix~\ref{sec:pseudocode-rect} and
Algorithm~\ref{alg:brute-force-nxnynz} to find   the 3719 states
$(n_x, n_y, n_z)$ that have $\kappa_{n_xn_yn_z}\leq 20\pi$ for 
 $\xi_x=\xi_y=1$ (set
$\kappa_{\max}=20\pi$ in the algorithm).   Generate plots similar to
Figs.~\ref{fig:nxnynz} and \ref{fig:Nparallelepiped},
including smooth curves corresponding to the inclusion of various
terms from the asymptotic expression for
$\overline{N}_{\rm rect}(\kappa)$ [see
Eq.~(\ref{eq:N-rect-four-terms})].  As a first step,    
check that setting $\kappa_{\max} = 4\pi$ yields the
same 17 states described in Appendix~\ref{sec:pseudocode-rect}.  If
you  use {\tt Mathematica}, a plot similar to
Fig.~\ref{fig:nxnynz} can be generated using {\tt
  ListPointPlot3D[\ldots]}.

\bigskip
\noindent
{\bf Project 1.2 -- Quasi-one-dimensional geometry}

\noindent
Use your code from Project 1.1 to find $N(\kappa)$ as a function of
$\kappa$ for   $\xi_x=50$ and $\xi_y=1$, setting
$\kappa_{\max} = 12\pi$.  In this case the $x$
dimension of the box is fifty times larger than the $y$ and $z$
dimensions, so the box is quasi-one-dimensional.  Plot $N(\kappa)$ as
as in Fig.~\ref{fig:Nparallelepiped}.  You
should see prominent cusps near $\kappa \approx 26$ and
$\kappa \approx 33$.

\bigskip
\noindent
{\bf Problem 1.2 -- Analysis of a quasi-one-dimensional geometry}

(a) Use Eq.~(\ref{eq:kappanxnynz}) to derive an 
expression for $N(\kappa)$ in the region
between $\kappa \approx 16$ and the cusp at $\kappa \approx 26$ in
Project~1.2.  Hint: In this region, $n_y=n_z=1$, so $N=n_x$.
Superimpose your theoretical expression on the plot of the data
obtained in Project~1.2.

(b) Use Eq.~(\ref{eq:kappanxnynz}) to calculate $\kappa$ and $N$ for
the cusp near $\kappa \approx 26$.  The cusp occurs when
$n_y$ and $n_z$ begin to take on values other than 1.  (The idea
is to calculate the location of the cusp, which will  be
close to $\kappa = 26$.)

(c) (More difficult) Use Eq.~(\ref{eq:kappanxnynz}) to determine
$N(\kappa)$ as a function of $\kappa$
between the cusps near $\kappa \approx 26$ and $\kappa \approx 33$.
Superimpose your result on the plot of the data obtained in Project~
1.2.

(d) Use Eq.~(\ref{eq:kappanxnynz}) to calculate $\kappa$ for the cusp
near $\kappa \approx 33$.  What is the reason for the cusp in this
case?

\bigskip
\noindent
{\bf Project 1.3 -- Linear coefficient for the quasi-one-dimensional geometry}

\noindent
Calculate $N(\kappa)$ for $\xi_x=50$ and $\xi_y=1$ (as in
Project~ 1.2)  to at least
$\kappa_{\max} = 40\pi$.  Subtract the volume and
area terms from $N(\kappa)$ to obtain $\Delta N$ [see
Eq.~(\ref{eq:DeltaN-rect})].  Plot $\Delta N$ as a function of
$\kappa$ and  do a linear fit.  Compare your result
to the  expression for the linear coefficient in
Eq.~(\ref{eq:f1rect}) [that is, compare your slope to
$f_{\rm rect}^{(1)}(50,1)$].  The two results should
agree to within 5\%.  In principle you should use the improved
prescription for determining $N(\kappa)$ here (see the discussion at
the end of Appendix~\ref{sec:pseudocode-rect}), although  
it does not make  much difference in this  case.

\bigskip
\noindent
{\bf Project 1.4 -- Quasi-two-dimensional geometry}

\noindent
This project can be done immediately after Project~1.1, although it
would be helpful to have read Project~1.2 first.  Use your code from
Project~1.1 to find $N(\kappa)$ as a function of $\kappa$ for  
$\xi_x=1/50$ and $\xi_y=1$, setting
$\kappa_{\max} = 30\pi$.  In this case the $x$
dimension is one fiftieth the length of the $y$ and $z$ dimensions, so
the box is quasi-two-dimensional.  Plot $N(\kappa)$ as a function of
$\kappa$ and superimpose the theoretical asymptotic expression.  You should see a
cusp near $\kappa \approx 85$.

\bigskip
\noindent
{\bf Problem 1.3 -- Analysis of the quasi-two-dimensional case (difficult)}

(a) Calculate the approximate location of the cusp near
$\kappa \approx 85$, and determine $N(\kappa)$ for $\kappa$ between $\kappa\approx 43$ and $\kappa \approx 85$.  You will probably
need to do a continuum approximation for $n_y$ and $n_z$.  Superimpose
your plot on the exact result obtained in Project~1.4.

(b) Determine an approximate expression for $N(\kappa)$ for $\kappa$
between $\kappa \approx 85$ and the next cusp.  Superimpose your result
on the exact result.

(c) Generalize your result to find an approximate expression for $N(\kappa)$
 between any two cusps.  Use your result to
recover the volume term in Eq.~(\ref{eq:N-rect-four-terms}),
showing that this quasi-two-dimensional example  approaches the
usual result for large $\kappa$.

\subsection{Sphere}

\noindent
{\bf Problem 2.1 -- Theoretical expressions}

\noindent
Start from Eqs.~(\ref{eq:kappadefn}) and (\ref{eq:knlm}) and derive
Eqs.~(\ref{eq:kappanlm}) and (\ref{eq:N-sphere-four-terms}).

\bigskip
\noindent
{\bf Project 2.1 -- $\bm{N(\kappa)}$}

\bigskip
\noindent
Modify your code from Project 1.1 and follow the approach  in
Appendix~\ref{sec:spherical-cylindrical-algorithms} to find  
the 3818 states $(n,m,\ell)$ that have $\kappa_{nlm}\leq 20\pi$ for the
spherical case.  
Generate plots similar to Figs.~\ref{fig:nml} and \ref{fig:Nk-sphere}, including smooth curves corresponding to the inclusion
of various terms from the asymptotic expression for
$\overline{N}_{\rm sph}(\kappa)$,
Eq.~(\ref{eq:N-sphere-four-terms}).

\bigskip
\noindent
{\bf Project 2.2 -- An efficient algorithm}

\noindent
Implement a much more efficient algorithm for the
spherical case.

(a) Use the algorithm outlined in
Appendix~\ref{sec:efficient-spherical-one-value} to determine the number
of states with $\kappa_{n\ell m}\leq 20\pi$.  Your answer should agree
with what you found in Project~2.1.  Also, determine the largest
eigenvalue $\kappa_{n \ell m}$ that is less than or equal to $20\pi$.  You
will probably find that your algorithm runs much more quickly than
your previous one.

(b) Repeat for $\kappa_{\max} = 500$.  Calculate the
percentage difference between your value for $N$ and the theoretical
expression in Eq.~(\ref{eq:N-sphere-four-terms}).  (Run your code from Project~2.1 for  
$\kappa_{\max} = 500$ so that you can compare the
run times.)

\bigskip
\noindent
{\bf Project 2.3 -- Numerical determination of the linear coefficient (difficult)}

\noindent
The goal of  this project, which extends Project~2.2,  is to
determine the linear coefficient in Eq.~(\ref{eq:N-sphere-four-terms})
numerically.  A complicating factor in the spherical case is that it is necessary
 to go to large values of $\kappa$   to obtain a
reasonable result.  Rather than calculating $\Delta N$ for all values
of $\kappa$ over some large range and performing a linear fit (as in
Fig.~\ref{fig:DeltaNparallelepiped}),   
calculate $\Delta N$ over a sequence of small intervals (for example,
$\kappa \in(20, 30)$, $(120, 130)$,\ldots.)  Then consolidate
the data in those ranges by computing averages, and do a linear
fit to those points.

(a) Implement the algorithm described in
Appendix~\ref{sec:efficient-spherical-many-values} for  
$\kappa_{\min}=100$ and
$\kappa_{\max}=120$.  Average the values as
described there to determine $\overline{\kappa}$ and
$\overline{\Delta N}$ (along with its uncertainty).  Embed
your algorithm in a function so that it can be called for any value
of $\kappa_{\min}$ and
$\kappa_{\max}$.

(b) Determine $\overline{\kappa}$ and $\overline{\Delta N}$ (along
with its uncertainty) for $\kappa \in(20, 30)$, $\kappa\in(120, 130)$,\ldots,
$\kappa\in(420, 430)$,\ldots.  Plot your ($\overline{\kappa}$,
$\overline{\Delta N}$) pairs, including error bars in the vertical
direction.  The data should look reasonably linear if the
error bars are taken into account.

(c) Do a weighted linear fit to your data from part~(b) taking
into account the uncertainties of the $\overline{\Delta N}$
values.~\cite{bevington}  Superimpose the fit on your plot from part~
(b).  Compare the slope that you obtain to the linear coefficient in
Eq.~(\ref{eq:N-sphere-four-terms}).

\subsection{Cylinder}

\noindent
{\bf Problem 3.1 -- Theoretical expressions}

(a) Start from Eq.~(\ref{eq:knmnz}) and
 the definitions  in
Eqs.~(\ref{eq:kappadefn}) and (\ref{eq:xia}) and derive
Eqs.~(\ref{eq:kappanmnz}) and (\ref{eq:N-cyl-four-terms}).

(b) Determine the value of $\xi_a$ that minimizes
$f_{\rm cyl}^{(2)}(\xi_a)$.  (This
value of $\xi_a$ minimizes the surface area of the cylinder for a
given volume.)

\bigskip
\noindent
{\bf Project 3.1 -- $\bm{N(\kappa)}$}

\noindent
Modify your code from Project~1.1 and follow the approach  in
Appendix~\ref{sec:spherical-cylindrical-algorithms} to find  
the 3761 states $(n_z,m,n)$ that have $\kappa_{nmn_z}\leq 20\pi$ for the
cylindrical case, with $\xi_a=1/3$.  Generate plots similar to Figs.~\ref{fig:Nk-sphere}  and
\ref{fig:nzmn}, including smooth curves
corresponding to the inclusion of various terms from the asymptotic
expression for $\overline{N}_{\rm cyl}(\kappa)$,
Eq.~(\ref{eq:N-cyl-four-terms}).

\bigskip
\noindent
{\bf Project 3.2 -- Linear coefficient (difficult)}

\noindent
Calculate $N(\kappa)$ for $\xi_a=1/3$ (as in Project~
3.1).  Reasonable results for the linear coefficient can be obtained
by choosing   $\kappa_{\max} = 20\pi$; better
agreement will be obtained by choosing $\kappa_{\max} = 150$, but the computation time will
be  long.  Subtract the volume and area terms in
Eq.~(\ref{eq:N-cyl-four-terms}) from $N(\kappa)$ to obtain $\Delta N$.
Plot $\Delta N$ as a function of $\kappa$ and  do a linear fit to
this data.  Compare your result to the  expression for the
linear coefficient in Eq.~(\ref{eq:f1cyl}) [that is, compare your slope
to $f_{\rm cyl}^{(1)}(1/3)$].  In principle, you should
use the improved prescription for determining $N(\kappa)$ here
(see the discussion at the end of Appendix~\ref{sec:pseudocode-rect}),
although in practice it does not make  much difference in this case.

\bigskip
\noindent
{\bf Project 3.3 -- A quantum wire}

\noindent

Use your code
from Project~3.1 to find $N(\kappa)$  for 
 $\xi_a=1/200$ with $\kappa_{\max} = 50$.
The length of the cylinder is 200 times larger than its
radius, so the cylinder is quasi-one-dimensional.  Plot $N(\kappa)$  and superimpose smooth curves
corresponding to the inclusion of various terms from the asymptotic 
expression, Eq.~(\ref{eq:N-cyl-four-terms}).  You should see prominent
cusps near $\kappa \approx 33$ and $\kappa \approx 44$.

\bigskip
\noindent
{\bf Problem 3.2 -- Analysis of a quantum wire}

\noindent
This problem is related to Project~3.3.  Analytically determine the
locations of the cusps near $\kappa \approx 33$ and
$\kappa \approx 44$.  (See Problem~1.2 for  help getting started.)

\bigskip
\noindent
{\bf Project 3.4 -- A thin disk}

\noindent
Use your code from Project~3.1 to find $N(\kappa)$  for   $\xi_a=15$, setting
$\kappa_{\max} = 60$.  The length
of the cylinder is 15 times smaller than its radius,
so the cylinder is quasi-two-dimensional.  You should see a
cusp near $\kappa \approx 56$.

\bigskip
\noindent
{\bf Problem 3.3 -- Analysis of a thin disk (very difficult)}

\noindent This problem is related to
Project~3.4.  Analytically determine the location of the cusp near
$\kappa \approx 56$ and obtain  an analytical expression for
$N(\kappa)$ for $\kappa$ between $\kappa \approx 28$ and $\kappa \approx 56$.  (To calculate $N(\kappa)$
you will need to follow an approach similar to that  in
Appendix~\ref{sec:derivation-cylinder}.)

\section{Discussion and Concluding Remarks}
\label{sec:discussion-conclusions}

In this work we presented algorithms to compute the cumulative state
number, $N(k)$, for a rectangular box, a sphere and a cylinder.  In
each case $N(k)$ is discrete and may be computed by summing a set of
points on a lattice in three dimensions.  Through projects and
problems described here, students can calculate $N(k)$ and compare
their exact results with various terms in the asymptotic expression.
As noted in Sec.~\ref{sec:intro} $N(k)$ and its derivative, the
density of states, have many applications in a wide range of physical
contexts.  Our analysis has been in the context of non-relativistic
quantum mechanics, but there are also classical applications (in
acoustics) and important relativistic applications, including
blackbody radiation and the Casimir
effect.~\cite{baltes-hilf,reudler,casimir}

We emphasized the analysis of quasi-one- and
two-dimensional bounding domains, for which $N(k)$ shows a mixture of
discrete and continuum-like behavior due to quantum numbers
associated with the small and large dimensions, respectively.  Plots
of $N(k)$ contain cusps that indicate changes in the quantum numbers
associated with the small dimensions.  Students can derive analytical
expressions that describe the regions between the cusps; in the
quasi-two-dimensional cases, the large $k$ behavior of these
expressions reduces to the leading term in the asymptotic expression,
verifying Weyl's theorem.  Nanostructures can be fabricated that
approximate one- and two-dimensional systems.
Reference~\onlinecite{quantum-ring} contains an experimental study and
analysis of the quasi-one-dimensional density of states in a quantum
ring.  Although the situation there is much more complicated than
those considered here, we note an interesting discussion in that paper
regarding the effect of the finite rim width compared to the ring
circumference.

There
are several ways in which the projects in this work can be extended or
generalized.  Here we list a few possibilities.

\begin{enumerate}
\item {\bf Two dimensions.}  The    asymptotic expression for
  the average cumulative state number is  well known in two
  dimensions, so we can consider, for example, rectangular and
  circular domains and analyze the agreement between the exact and
  asymptotic results.  See 
  Ref.~\onlinecite{baltes-hilf}, pp.~59--63  as a starting point.
 
\item {\bf Electromagnetic problem.}  We have implicitly 
  considered what is called the scalar version of the problem,
  but we can also consider the electromagnetic version of the
  problem, which is important for blackbody radiation.  In this case
  we need to consider correlated electric and magnetic fields, and
  their corresponding boundary conditions.  See  Ref.~\onlinecite{baltes-hilf} and references therein.
\item {\bf Bubble/thick-shell example.}  The theoretical discussion  in Ref.~\onlinecite{bereta-et-al}  can be used to solve for the
  eigenvalues of wave functions inside a spherical shell with Dirichlet
  boundary conditions.  Reference~\onlinecite{baltes-hilf} and references
  therein contain useful information for determining the asymptotic
  expression for the averaged cumulative state number.
  Reference~\onlinecite{gutierrez-yanez} could also be helpful (but  the results in Ref.~\onlinecite{gutierrez-yanez} are written in
  terms of a time variable; see   Ref.~\onlinecite{baltes-hilf}, pp.~30--31.)
\item {\bf A complementary method of determining $\bm{N}$ in the
    spherical case.}  We can use Eq.~(\ref{eq:nmaxl-continuum}) 
  with the factor of $1/4$ replaced by $\chi/\pi$ to determine $n$ versus
  $\ell$ (for integer $\ell$ values) for the spherical case.  Rounding the
  $n$ values down to integers   gives an accurate boundary
  that can then be used to determine the exact number of states
  without  determining the zeros of Bessel functions (which is
  computationally expensive).  The trick is that  $\chi$  must be
  determined accurately, which can be done by computing a
  line integral in the complex plane, as described in
  Ref.~\onlinecite{watson}.
\end{enumerate}

\acknowledgments
KK thanks Taylor University for its support during his sabbatical.
The authors thank Daniel Schroeder for helpful feedback.  Solutions to   the problems and sample
code for the projects are available upon
request by sending an email to KK.

\appendix

\renewcommand{\theequation}{\Alph{section}.\arabic{equation}}

\section{Summary of Energy Eigenfunctions}
\label{sec:app-energy-eigenfns}

We list the energy eigenfunctions and eigenvalues
(the solutions to Eq.~(\ref{eq:TISE}) with Dirichlet boundary
conditions) for the three geometries  considered in this work.
The rectangular case is a straightforward generalization of the
one-dimensional version.  Results for the spherical and cylindrical
cases may be found, for example, in Refs.~\onlinecite{mulhall-moelter,
  lambert}.

\subsection{Rectangular parallelepiped}
\label{sec:app-rect-energy-eigenfns}

The energy eigenfunctions in the rectangular domain
$0\leq x\leq L_x$, $0\leq y\leq L_y$ and $0\leq z\leq L_z$ are given
by
\begin{equation}
  \psi_{n_xn_yn_z}(x,y,z) = A_{n_xn_yn_z} \sin\left(\frac{n_x\pi x}{L_x}\right)
  \sin\left(\frac{n_y\pi y}{L_y}\right)\sin\left(\frac{n_z\pi z}{L_z}\right) \,,
\end{equation}
where $ A_{n_xn_yn_z}$ is a normalization constant and   
the integers $n_x$, $n_y$ and $n_z$ can independently take on the
values $1, 2, 3,\ldots$.  The corresponding energy eigenvalues are
given by
\begin{equation}
  \epsilon_{n_xn_yn_z} = \frac{\hbar^2\pi^2}{2\mu}\left(
  \frac{n_x^2}{L_x^2}+\frac{n_y^2}{L_y^2}+\frac{n_z^2}{L_z^2}\right)\, .
  \label{eq:epsnxnynz}
\end{equation}

\subsection{Sphere}
\label{sec:app-sphere-energy-eigenfns}

The energy eigenfunctions inside the spherical domain
$r\leq a$ are given by
\begin{equation}
  \psi_{n \ell m}(r,\theta,\phi) = A_{n \ell m} Y_{\ell m}(\theta,\phi) j_\ell\left(\beta_{\ell,n}r/a\right) \,,
\end{equation}
where $A_{n\ell m}$ is a normalization constant, $Y_{\ell m}(\theta,\phi)$ is
a spherical harmonic and $j_\ell\left(\beta_{\ell,n}r/a\right)$ is the
spherical Bessel function of order $\ell$.  $\beta_{\ell,n}$ is the $n$th
zero of $j_\ell$.  The 
quantum numbers are given by $n=1,2,3,\ldots$, $\ell=0,1,2,\ldots$ and
$m=-\ell, -\ell+1,\ldots,\ell$.  The energy eigenvalue corresponding to the
state $\psi_{n\ell m}$ is
$
\epsilon_{n\ell m} = \hbar^2\beta_{\ell,n}^2/\left(2\mu a^2\right)  
$.
The energy eigenstates are $(2\ell+1)$-fold degenerate, because
the corresponding energy eigenvalues do not depend on $m$.

\subsection{Circular cylinder}
\label{sec:app-cylinder-energy-eigenfns}

The energy eigenfunctions inside the cylindrical domain bounded by
$r\leq a$ and $0\leq z \leq L$ are given in cylindrical coordinates by
\begin{equation}
  \psi_{nmn_z}(r,\phi,z) = A_{nmn_z} J_m\left(\zeta_{m,n}r/a\right)e^{im\phi}
  \sin\left(\frac{n_z\pi z}{L}\right)\,,
\end{equation}
where $A_{nmn_z}$ is a normalization constant,
$J_m(\zeta_{m,n}r/a)$ is the Bessel function of order $m$,
and $\zeta_{m,n}$ is the $n$th zero of $J_m$.  The quantum numbers can take on the integer
values $n=1,2,3,\ldots$, $m=0, \pm1, \pm2,\ldots$ and
$n_z=1,2,3,\ldots$ and the energy eigenvalues are
\begin{equation}
  \epsilon_{nmn_z} = \frac{\hbar^2}{2\mu}
  \left(\frac{\zeta_{m,n}^2}{a^2}+\frac{\pi^2n_z^2}{L^2}\right) \, .
\end{equation}
Note that $J_{-m}(x) = (-1)^mJ_m(x)$ for
$m=1,2,3\ldots$,~\cite{besselref} meaning that $J_{-m}$ and $J_m$ have
zeros at the same locations.  Thus, the energy eigenvalues have a
two-fold degeneracy for each $m\neq 0$.

\section{Pseudocode for determining $\bm{N}$}
\label{sec:pseudocode}

\subsection{The rectangular case}
\label{sec:pseudocode-rect}

$N(\kappa)$ is the number of states with
$\kappa^2_{n_xn_yn_z}\leq \kappa^2$ [see the definition of
$\kappa^2_{n_xn_yn_z}$ in Eq.~(\ref{eq:kappanxnynz})].  Our goal is to find  
$N(\kappa_{\max})$ and $N(\kappa)$ for all
$\kappa \leq \kappa_{\max}$.  We need to be careful not to
miss contributing states.  In 
Algorithm~\ref{alg:brute-force-nxnynz} we use several embedded {\tt
  While} loops and are careful to set appropriate exit conditions so
that the program does not become stuck in an infinite loop.\cite{footnote8}
We can visualize the algorithm with the aid of Fig.~\ref{fig:nxnynz}.
We start at the point $(n_x,n_y,n_z)=(1,1,1)$ and increase $n_z$ until
we reach the maximum value allowed by
$\kappa_{\max}$.  We then move to
$(n_x,n_y,n_z)=(1,2,1)$ and do the same until the entire back ``wall''
(with $n_x=1$) has been filled in.  At that point we increment $n_x$
and continue in the same manner until all of the states have been
found.   We have implemented the algorithm  in {\tt Mathematica} and {\tt Python}.
In Mathematica, the sorting and splitting can be done in a
single step using {\tt Split[Sort[dataArray, \#1<\#2
  \&]]}.\cite{footnote9}
We set $\kappa_{\max}=4\pi$ and $\xi_x=\xi_y=1$ and
obtain the following results for several of the arrays  in the pseudocode,
\renewcommand{\arraystretch}{1}
\begin{eqnarray}
\begin{array}{ccccc}
  \left(\begin{array}{ccc}
          1 & 1 & 1\\
          1 & 1	& 2\\
          1 & 1 & 3\\
          1 & 2 & 1\\
          1 & 2 & 2\\
             1 & 2 & 3\\
             1 & 3 & 1\\
           1 & 3 & 2\\
             2 & 1 & 1\\
            & \vdots & \\ \end{array}
          \right) &
  \left(\begin{array}{c}
          3\pi^2 \\6\pi^2\\11\pi^2\\6\pi^2\\9\pi^2\\14\pi^2\\11\pi^2\\14\pi^2\\6\pi^2\\ \vdots \\
        \end{array}
  \right) &
  \left(\begin{array}{c}
          3\pi^2 \\6\pi^2\\6\pi^2\\6\pi^2\\9\pi^2\\9\pi^2\\9\pi^2\\11\pi^2\\11\pi^2\\ \vdots \\
          \end{array}
  \right) &
  \left(\begin{array}{c}
             \{3\pi^2\} \\ \{6\pi^2, 6\pi^2, 6\pi^2\} \\ \{9\pi^2, 9\pi^2, 9\pi^2\} \\ \{11\pi^2, 11\pi^2, 11\pi^2\} \\ \{12\pi^2\}\\
             \{14\pi^2, 14\pi^2, 14\pi^2, \\~~14\pi^2, 14\pi^2, 14\pi^2\} \\
          \end{array}
  \right) &
          \left(\begin{array}{cc}
          5.4414 &	1 \\
          7.6953 &	4 \\
          9.42478 &	7 \\
          10.4195 &	10 \\
          10.8828 &	11 \\
          11.7548&	17 \\
        \end{array}\right) \,, \\
  & & & \\
  \mbox{\tt array \#1} & \mbox{\tt array \#2} & \mbox{\tt array \#3} & \mbox{\tt array \#4} & \mbox{\tt array \#5}\\
\end{array}
\end{eqnarray}
where {\tt array \#1} contains the triplets $(n_x,n_y,n_z)$ (which
can be used to generate a plot such as Fig.~\ref{fig:nxnynz}), {\tt
  array \#2} ({\tt \#3}) contains the unsorted (sorted) $\kappa_{n_xn_yn_z}^2$ values,
{\tt array \#4} contains the ``sorted and split'' values of $\kappa^2$
(note the degeneracies), and {\tt array \#5} contains
$\left(\kappa ,N(\kappa)\right)$ pairs (which can be used to
generate a plot such as Fig.~\ref{fig:Nparallelepiped}).  Note that
$N(\kappa)$ is  the number of states up to and including a
particular value of $\kappa$.

\renewcommand{\arraystretch}{2}
To determine the coefficient of the linear term in
Eq.~(\ref{eq:N-rect-four-terms}), it is important to use the
improved prescription for calculating $N$ that is defined in
Eq.~(\ref{eq:Nkdefn-improved}), instead of the original one defined in
Eq.~(\ref{eq:Nkdefn}).  Algorithm~\ref{alg:brute-force-nxnynz}
 uses the unimproved prescription, but it is
straightforward to fix this by  averaging successive $N$ values
in the array containing the $\kappa$ and $N$ values.  For example, for
the above data set, averaging yields:
\renewcommand{\arraystretch}{1}
\begin{eqnarray}
  \left(\begin{array}{cc}
          5.4414 &	1 \\
          7.6953 &	4 \\
          9.42478 &	7 \\
          10.4195 &	10 \\
          10.8828 &	11 \\
          11.7548&	17 \\
        \end{array}\right)
  \to
  \left(\begin{array}{cc}
          5.4414 &	0.5 \\
          7.6953 &	2.5 \\
          9.42478 &	5.5 \\
          10.4195 &	8.5\\
          10.8828 &	10.5 \\
          11.7548&	14\\
        \end{array}\right)  \,.
\end{eqnarray}
Subtracting the volume and area terms
from the $N(\kappa)$ values in the second array yields $\Delta N$ [see
Eq.~(\ref{eq:DeltaN-rect})], which can then be used to produce a
plot such as Fig.~\ref{fig:DeltaNparallelepiped}.
\renewcommand{\arraystretch}{2}

\begin{singlespace}
\begin{algorithm}[H]
\begin{verbatim}
set kappaMax, xix and xiy;
initialize array #1 (stores (nx,ny,nz) satisfying kappaSq<=kappaMax^2);
initialize array #2 (stores kappaSq values);
nx=1;
nxOK=true;
while nxOK do:
|  ny=1;
|  nyOK=true;
|  while nyOK do:
|  |  nz=1;
|  |  nzOK=true;
|  |  while nzOK do:
|  |  |  kappaSq=pi^2(nx^2(xiy/xix^2)^(2/3)+...);
|  |  |  if kappaSq>kappaMax^2 then:
|  |  |  |  nzOK=false;
|  |  |  |  if nz==1 then:
|  |  |  |  |  nyOK=false;
|  |  |  |  |  if ny==1 then:
|  |  |  |  |  |  nxOK=false;
|  |  |  |  |  else:
|  |  |  |  |  |  increment nx;
|  |  |  |  |  end;
|  |  |  |  else:
|  |  |  |  |  increment ny;
|  |  |  |  end;
|  |  |  else:
|  |  |  |  append (nx,ny,nz) to array #1;
|  |  |  |  append kappaSq to array #2;
|  |  |  |  increment nz;
|  |  |  end;
|  |  end;
|  end;
end;
sort array #2 by increasing values of kappaSq; store as array #3;
split array #3 into sets of degenerate values; store as array #4;
initialize array #5 (stores (kappa,N) pairs);
loop through array #4; count states and append results to array #5;
\end{verbatim}
  \caption{An algorithm to find all states in a rectangular
    parallelepiped that have $\kappa_{n_xn_yn_z}$ less than or equal
    to    $\kappa_{\max}$.  The
    states are sorted by increasing $\kappa$ and then degenerate
    states are collected into groups.  Array \#1 [which stores
    $(n_x,n_y, n_z)$ values] can be used to generate a figure such as
    Fig.~\ref{fig:nxnynz}, but is otherwise not needed.  Array \#5
    [which contains
    $\left(\kappa,N(\kappa)\right)$ pairs] can
    be used to generate figures such as
    Fig.~\ref{fig:Nparallelepiped}.}
  \label{alg:brute-force-nxnynz}
\end{algorithm}
\end{singlespace}

\subsection{Algorithms for the spherical and cylindrical geometries}
\label{sec:spherical-cylindrical-algorithms}

The approach  in
Algorithm~\ref{alg:brute-force-nxnynz} can be adapted for the spherical
and cylindrical cases with some small changes. For the spherical case, we replace $\kappa_{n_xn_yn_z}$ by
$\kappa_{n \ell m}$, and in the {\tt While} loops, we replace $n_x$ by $n$
and $n_y$ by $\ell$ [see Eq.~(\ref{eq:kappanlm}) and Fig.~\ref{fig:nml}].
The inner {\tt While} loop in Algorithm~\ref{alg:brute-force-nxnynz}
is not  necessary and can be replaced by a {\tt Do} loop for $m$
ranging from $-\ell$ to $\ell$.  (If   only states are counted, in contrast to generating a plot such as Fig.~\ref{fig:nml}, we can
 include a degeneracy factor.)  Another small change is that
$\ell$ starts at $0$ ($n_y$  starts at $1$).  {\tt Mathematica}
has a function that returns the $n$th zero of the regular Bessel
function.  Because
$j_\ell(x)\propto x^{-1/2}J_{\ell+1/2}(x)$,~\cite{besselref} finding the
$n$th zero of $j_\ell$ is equivalent to finding the $n$th zero of
$J_{\ell+1/2}$.  In {\tt Mathematica} $\beta_{\ell,n} = \mbox{\tt
  BesselJZero[$\ell+1/2$,$n$]}$.  For a useful algorithm in {\tt Python}
see the {\tt SciPy Cookbook}.~\cite{scipy-cookbook}

For the cylindrical case, we replace $\kappa_{n_xn_yn_z}$ by
$\kappa_{nmn_z}$, and in the {\tt While} loops, we replace $n_x$ by
$n_z$, $n_y$ by $m$, and $n_z$ by $n$, noting that $n_z$ starts at
$1$, $m$ is symmetrically distributed about $0$ and $n$ starts at $1$
[see Eq.~(\ref{eq:kappanmnz}) and Fig.~\ref{fig:nzmn}].  For $m$, we
can start at zero and increment through positive values, remembering
to include an extra state $-m$ for every state $m$ when $m>0$.

\subsection{More efficient algorithms for  spherical geometry}

The straightforward, brute force algorithms we have described  become
computationally expensive for large $\kappa$.  Furthermore, for the
spherical case, the difference between $N(\kappa)$ and the
averaged approximation in Eq.~(\ref{eq:N-sphere-four-terms}) is
highly oscillatory.  If we wish to perform a  fit to
determine the linear coefficient numerically for the spherical case,
we need to go to  large values of $\kappa$.  Fortunately, we
can employ a much more efficient algorithm if we are  interested
only in determining $N$ for a particular value of $\kappa$ or for a small
range of $\kappa$ values.

\subsubsection{$N$ for one value of $\kappa$}

\label{sec:efficient-spherical-one-value}


The pseudocode in Algorithm~\ref{alg:efficient-spherical}    
efficiently   determines $N$ for a given value
$\kappa_{\max}$.  This algorithm is based on three
observations, the first two of which may be understood by examining
Fig.~\ref{fig:nml}:
\begin{enumerate}
\item For a given value of $\kappa_{\max}$, the
  maximum value of $\ell$ is a monotonically decreasing function of $n$.
  (In Fig.~\ref{fig:nml}, the triangles become smaller as $n$
  gets larger.)
\item Once the maximum value of $\ell$ is known for a given value of $n$,
  the number of points in the corresponding triangle in
  $(n,m,\ell)$-space can be calculated immediately and is given by
\begin{equation}
  \sum_{\ell=0}^{\ell_{\max}}(2\ell+1) = \left(\ell_{\max}+1\right)^2 \;,
  \label{eq:num-states-triangle}
\end{equation}
  where $\ell_{\max}$ is the maximum value of $\ell$ for a
  given value of $n$.
\item The zeros of the spherical Bessel functions, $\beta_{\ell,n}$,
  satisfy the inequality
  $\beta_{\ell,n}\geq \ell+\frac{1}{2}$.~\cite{watson}  For $n=1$, 
  $\ell_{\max}$ is less than
  $\left(3/(4\pi)\right)^{1/3}\kappa_{\max}$ [see
  Eq.~(\ref{eq:kappanlm})].
\end{enumerate}
The idea of the algorithm is to start at $n=1$ and set $\ell$
equal to an integer slightly greater than
$\left(3/(4\pi)\right)^{1/3}\kappa_{\max}$.  To
determine $\ell_{\max}$, we decrease $\ell$ in unit
steps until $\beta_{\ell,1}$ is less than or equal to
$\left(3/(4\pi)\right)^{1/3}\kappa_{\max}$.  Having
found $\ell_{\max}$ for $n=1$, we add up the number
of states for this triangle [see
Eq.~(\ref{eq:num-states-triangle})] and proceed to $n=2$, decreasing
$\ell$ as necessary until $\beta_{\ell,2}$ is less than or equal to
$\left(3/(4\pi)\right)^{1/3}\kappa_{\max}$.

\bigskip
\begin{singlespace}
\begin{algorithm}[H]
\begin{verbatim}
set kappaMax;
initialize N to zero, n to 1, and kappaLargest to zero;
lMax=Ceiling((3/(4pi))^(1/3)kappaMax);
while lMax>-1 do:
|  solutionFound=false;
|  while ((not solutionFound) and (lMax>-1)) do:
|  |  betalMaxn=nth zero of spherical Bessel function;
|  |  kappa=(4pi/3)^(1/3)betalMaxn;
|  |  if kappa<=kappaMax then:
|  |  |  if kappa>kappaLargest then:
|  |  |  |  kappaLargest=kappa;
|  |  |  end;
|  |  |  increment N by (lMax+1)^2;
|  |  |  solutionFound=true;
|  |  else:
|  |  |  decrement lMax;
|  |  end;
|  end;
|  increment n;
end;
\end{verbatim}
\caption{An efficient algorithm to determine $N$, the number of states
    in a sphere,  for $\kappa_{n \ell m}\leq \kappa_{\max}$.
    The algorithm returns $N$ and the largest value
  of $\kappa_{n \ell m}$ encountered, $\kappa_{\rm largest}$.}
  \label{alg:efficient-spherical}
\end{algorithm}
\end{singlespace}

\subsubsection{$N$ for a range of $\kappa$ values}
\label{sec:efficient-spherical-many-values}

%
\begin{figure}[t]
\centering
  \resizebox{6.5in}{!}{\includegraphics*{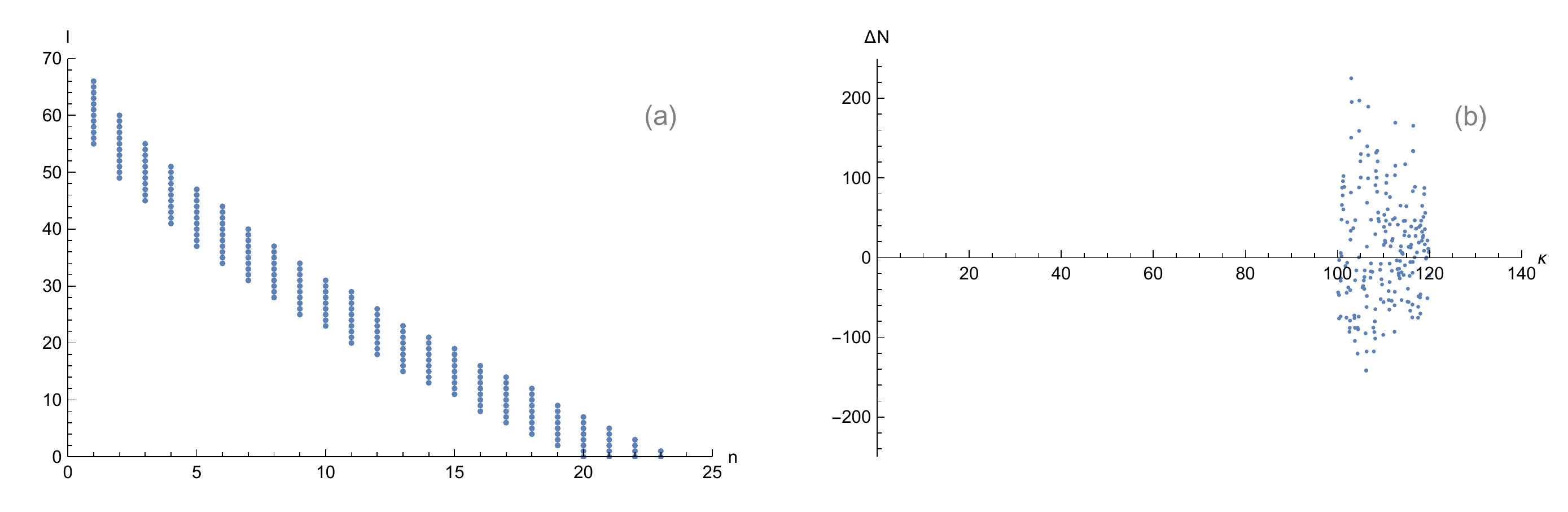}}
  \caption{(a) States with $100 \leq \kappa_{nlm} \leq 120$ for a
    spherical volume.  The $m$ direction has been suppressed, meaning
    that each dot represents $(2\ell+1)$ states.  (b) $\Delta N$ as a
    function of $\kappa$ for the states shown in (a).  To obtain
    $\Delta N$, we have subtracted the volume ($\sim\kappa^3$) and
    surface area ($\sim\kappa^2$) terms shown in
    Eq.~(\ref{eq:N-sphere-four-terms}).  For Project~2.3 we
    consolidate all of these points by averaging them, which yields
    $(\overline{\kappa},\overline{\Delta N}) \simeq (110.229,
    11.6\pm4.9)$.}
\label{fig:nlsphere}
\end{figure}
%

In Project~2.3 we   must  determine all
$(\kappa,N(\kappa))$ pairs for
$\kappa_{\min} \leq \kappa \leq
\kappa_{\max}$.
Algorithm~\ref{alg:efficient-spherical} can be adapted for this task
with some small adjustments.  Instead of immediately
incrementing $n$ once a solution has been found for some value of $\ell$,
we continue to decrement $\ell$ until $\kappa <
\kappa_{\min}$ (or $\ell$ becomes negative).  The values of $n$, $\ell$
and $\kappa$  can be stored in an array, which can then be
sorted in order of increasing $\kappa$ values.  Because $N$ is known for
the largest value of $\kappa$, the values of $N$ for the other
$\kappa$ values can be determined by traversing the sorted array
(backwards) and subtracting appropriate values of $(2\ell+1)$ from the
running value of $N$ at each step.

As an example, suppose we wish to find all of the $(\kappa, N(\kappa))$ pairs with
$100 \leq \kappa \leq 120$.  Figure~\ref{fig:nlsphere}(a) shows the corresponding
states in the $(n,\ell)$ plane (each point  represents $(2\ell+1)$
states, because we have suppressed the $m$ direction in the plot).  The largest
value of $\kappa$ in this range is  $\approx 119.933$, and
corresponds to $(n,\ell) = (15,19)$ and $N=27,745$.  The next largest value of $\kappa$
is  $\approx 119.846$ and the corresponding value of $N$ is thus
$27,745-(2\times19+1) = 27,706$.  Continuing in this way yields the
$(n,\ell,\kappa, N)$ array in the following:
\renewcommand{\arraystretch}{1}
\begin{eqnarray}
\begin{array}{cccccc}
  \left(\begin{array}{cccc}
          5 & 37 & 100.146 & 15,990\\
          9 & 25 & 100.342 & 16,041\\
          \vdots & \vdots & \vdots & \vdots\\
          6 & 44 & 119.555 & 27,548\\
          1 & 66 & 119.728 & 27,681\\
          18 & 12 & 119.846 & 27,706 \\
          15 & 19 & 119.933 & 27,745 \\ \end{array}
  \right)\!\! & \!\!\to \!\!& \!\!
  \left(\begin{array}{cc}
          100.146 & 15,952.5\\
          100.342 & 16,015.5\\
          \vdots & \vdots\\
          119.555 & 27,503.5\\
          119.728 & 27,614.5\\
          119.846 & 27,693.5 \\
          119.933 & 27,725.5 \\ \end{array}
  \right)\!\! & \!\!\to \!\!&\!\!
  \left(\begin{array}{cc}
          100.146 & -43.593\\
          100.342 & -76.382\\
          \vdots & \vdots\\
          119.555 & 21.482\\
          119.728 & 11.113\\
          119.846 & 6.858 \\
          119.933 & -22.426 \\ \end{array}
  \right)
  \\
  & & \\
  (n,\ell,\kappa, N) & & (\kappa, N_{\rm improved}) & & (\kappa, \Delta N)\\
 \end{array}
\end{eqnarray}
For Project~2.3 we need to manipulate the data in the $(n,\ell,\kappa, N)$ array
to determine average values for $\kappa$ and
$\Delta N$.  In the $(\kappa, N_{\rm improved})$ array we implement the
improved definition of $N$ [see Eq.~(\ref{eq:Nkdefn-improved}) and
also the discussion in Appendix~\ref{sec:pseudocode-rect}; note that
in this case we can  subtract appropriate values of
$(2\ell+1)/2$].  Then we subtract the volume and area terms in Eq.~(\ref{eq:N-sphere-four-terms}) to determine
$\Delta N$.  The result is shown in the $(\kappa, \Delta N)$ array above and is 
plotted in Fig.~\ref{fig:nlsphere}(b).  The $\Delta N$ values are
highly oscillatory.  Finally, we determine the mean values of the $(\kappa, \Delta N)$ array,
which yields
$(\overline{\kappa},\overline{\Delta N}) \simeq (110.229,
11.6\pm4.9)$.  To estimate the uncertainty in $\overline{\Delta N}$,
we have computed the standard deviation and divided by the square root
of 208 (which is the number of data points that are being averaged).
Note that the amplitude of the oscillations in $\Delta N$ is much
greater than the mean value.

\section{The volume and surface area terms for the spherical and cylindrical cases}
\label{sec:derivation-volume-surface}

In this appendix we provide a quasi-rigorous derivation of the volume
and surface area terms in Eq.~(\ref{eq:Navge-four-terms}) for a sphere
and cylinder.\cite{footnote10}
Reference~\onlinecite{lambert} contains similar calculations for the density
of states $dN(k)/dk$, but only of the volume term.\cite{footnote11}  Because we 
calculate both the volume and the surface area terms, we need to be
more careful  how the bounding surfaces are treated.  Also,
Ref.~\onlinecite{lambert} neglects the dependence on $\chi$ [see Eq.~(\ref{eq:bound})], but we find that this quantity must be treated  carefully
to obtain the correct expression for the surface area terms.

\subsection{Sphere}
\label{sec:derivation-volume-surface-sphere}

It is  nontrivial to determine the volume of the lattice
points bounded by a particular choice of $k$ in the large $k$ limit,
because we need to be able to determine the zeros of spherical Bessel
functions for (potentially)  large  $\ell$.  In particular, we need
to determine an expression that describes the top surface
in Fig.~\ref{fig:nml} for  large $k$  (the maximum value for
$\ell$ as a function of $n$, or vice versa).  It turns out that $\beta_{\ell,n}$, the
$n$th zero of the regular Bessel function of
order $\ell+\frac{1}{2}$,  satisfies the 
equality,~\cite{watson, watson2}
\begin{equation}
  \left(\ell+\frac{1}{2}\right)\left[\tan(\alpha) -\alpha\right] + \chi = n\pi
  \label{eq:bound}
\end{equation}
(except possibly for $n=1$), where
\begin{equation}
  \alpha \equiv \cos^{-1}\left(\frac{\ell+\frac{1}{2}}{\beta_{\ell,n}}\right) \; ,
  \label{eq:alpha}
\end{equation}
with $0\leq \alpha \leq \pi/2$; furthermore,
$\beta_{\ell,n}\geq \ell+\frac{1}{2}$.  The quantity $\chi$ is defined as the phase of
a particular integral along a contour in the complex plane (see Ref.~\onlinecite{watson}) and is such that
$
0 < \chi < \pi/4.
$.
For large zeros of Bessel functions, $\chi$ approaches
$\pi/4$; that is, for fixed $\ell$,~\cite{watson, footnote12}
$
 \lim_{\beta_{l,n}\to\infty} \chi = \frac{\pi}{4}
$.
For large $k$,  the relation between $n$ and $\ell$ on
the integration boundary (that is, when $\beta_{\ell,n}\simeq ka$) is given
approximately by
\begin{equation}
  n_{\rm sph} = \frac{\left(\ell+\frac{1}{2}\right)}{\pi}
    \left[\tan\left(\alpha_{\rm sph}\right)-\alpha_{\rm sph}\right] + \frac{1}{4}\; ,
  \label{eq:nmaxl-continuum}
\end{equation}
where
\begin{equation}
  \alpha_{\rm sph} \equiv\cos^{-1}\left(\frac{\ell+\frac{1}{2}}{ka}\right) \;.
  \label{eq:alpham}
\end{equation}

Let us consider the lower limit for the integration in the $n$
direction as well.  The lower limit for the sum over $n$ is
$n=1$.  However, when approximating the sum by an integral, we need to
be careful when all of the points lie exactly on the integration
boundary (see the discussion between Eqs.~(\ref{eq:A2ij}) and
(\ref{eq:kappadefn})).
Two options in this case are  to extend the volume integral to the
(unphysical) limit $n=1/2$, which effectively captures all of the
points having $n=1$, or  to keep the integration boundary at
$n=1$, but compute an area integral correction to make up for the fact
that only half of the points at $n=1$ are captured in this case.  Both
approaches give the same result to order $k^2$, so we choose the
former approach.

It remains to determine the extreme upper limit for the integral over
$\ell$ (when $n=1$).  To this end, we note the  asymptotic
expression,~\cite{abramowitz-stegun}
\begin{equation}
  \beta_{\ell,1} \sim \left(\ell+\frac{1}{2}\right) +1.8557571 \left(\ell+\frac{1}{2}\right)^{1/3}+\ldots \; .
  \label{eq:betal1}
\end{equation}
Thus, replacing $\beta_{\ell,1}$ by $ka$, we see that when $n=1$, the
maximum value of $\ell$ for large $ka$ is approximately $ka - 1/2$.  The
integral is then\cite{footnote13}
\begin{align}
  N_{\rm sph} (k) & =  \int_0^{ka-1/2}\left(2\ell+1\right)\int_{\frac{1}{2}}^{n_{\mbox{\tiny sph}}}dn \, d\ell
           \\
    & =  2 \! \int_0^{ka-1/2}
          \left(\ell+\frac{1}{2}\right)\left[ \frac{\left(\ell+\frac{1}{2}\right)}{\pi}
          \left[\tan\left(\alpha_{\rm sph}\right)-\alpha_{\rm sph}\right] +\frac{1}{4} - \frac{1}{2} \right] d\ell \; .
          \label{eq:Nvolsphere}
\end{align}
For the terms involving $\alpha_{\rm sph}$, we use
Eq.~(\ref{eq:alpham}) to perform a change of variables from $\ell$ to
$\alpha_{\rm sph}$ (noting that
$\alpha_{\rm sph}\sim\pi/2-1/(2ka)$ when $\ell= 0$ and
$ka$ is large).  The resulting integrations are then readily performed, and we
have
\begin{align}
  N_{\rm sph} (k) & =  \frac{2\left(ka\right)^3}{9\pi} -\frac{1}{4}\left(ka\right)^2 +\ldots \\
     & =  \frac{V k^3}{6 \pi^2} - \frac{Sk^2}{16\pi} + \ldots \; 
\end{align}
which agrees with the expression in Eq.~(\ref{eq:Navge-four-terms}).

\subsection{Circular cylinder}
\label{sec:derivation-cylinder}

The integral approximation of $N(k)$ is arguably more complicated in
this case than   in the previous two, although many of the
considerations are similar to those we encountered  for the
spherical case.  The integration boundary for $n$ in terms of $m$ and
$n_z$ is given by
\begin{equation}
  n_{\rm cyl} = \frac{m}{\pi}\left[\tan
    \left(\alpha_{\rm cyl}\right)-\alpha_{\rm cyl}\right] +\frac{1}{4}\; ,
  \label{eq:nmax}
\end{equation}
where
\begin{equation}
  \alpha_{\rm cyl} \equiv\cos^{-1}\left(\frac{m}{a\sqrt{k^2-\frac{\pi^2n_z^2}{L^2}}}\right) \;.
  \label{eq:alpham-cyl}
\end{equation}
The sum of states may  be approximated by
\begin{align}
  N_{\rm cyl} (k) & =  2\int_{\frac{1}{2}}^{n_{z,\mbox{\tiny max}}}\int_0^{m_{\mbox{\tiny bdy}}}\int_{\frac{1}{2}}^{n_{\mbox{\tiny cyl}}}dn \, dm \, dn_z
           \\
    & =  2\int_{\frac{1}{2}}^{n_{z,\mbox{\tiny max}}}\int_0^{m_{\mbox{\tiny bdy}}}
          \left[ \frac{m}{\pi}
          \left[\tan\left(\alpha_{\rm cyl}\right)-\alpha_{\rm cyl}\right] +\frac{1}{4} - \frac{1}{2} \right] dm \, dn_z \; ,
          \label{eq:Nvolcylinder}
\end{align}
where $m_{\rm bdy}$ represents the large $k$ limit of
the boundary for $m$ as a function of $n_z$ when $n=1$.  The endpoint
of the integration over $n_z$ is slightly less than $kL/\pi$ (the
strict upper limit for the sum is
$(kL/\pi)\sqrt{1-\zeta_{0,1}^2/(ka)^2}$).

To perform the integration over $m$, we use Eq.~(\ref{eq:alpham-cyl})
to change variables to $\alpha_{\rm cyl}$, noting that
$\alpha_{\rm cyl}\to \pi/2$ as $m\to 0$.  We denote the
limit of $\alpha_{\rm cyl}$ as
$m\to m_{\rm bdy}$ by
$\alpha_{\rm bdy}$, which satisfies the 
expression,
\begin{equation}
  \sin\left(\alpha_{\rm bdy}\right) - \alpha_{\rm bdy} \cos\left(\alpha_{\rm bdy}\right)
  = \frac{3\pi}{4}\frac{1}{a\sqrt{k^2-\frac{\pi^2n_z^2}{L^2}}} \; ,
  \label{eq:alphabdy}
\end{equation}
leading to
\begin{eqnarray}
  \!\!N_{\rm cyl} (k) \!&\simeq& \!\frac{k^3 a^2L}{6\pi} -\frac{k^2a^2}{8}\nonumber \\
    &&{} +  \!\!\! \frac{a^2}{4\pi}\int_{1/2}^{n_{z,\mbox{\tiny max}}}\!\left[k^2-\frac{\pi^2 n_z^2}{L^2}\right]\!\!
    \left[-2 \alpha_{\rm bdy} \sin^2\left(\alpha_{\rm bdy}\right)
      + \frac{5\pi\cos\left(\alpha_{\rm bdy}\right)}{2a\sqrt{k^2-\frac{\pi^2n_z^2}{L^2}}}\right]dn_z .
\end{eqnarray}
It turns out that $\alpha_{\rm bdy}$ is small when $k$
is large and $n_z$ is not too close to $kL/\pi$.  We expand the
left-hand side of Eq.~(\ref{eq:alphabdy}) for small
$\alpha_{\rm bdy}$ to find
\begin{equation}
  \alpha_{\rm bdy}^3 \simeq \frac{9\pi}{4}\frac{1}{a\sqrt{k^2-\frac{\pi^2n_z^2}{L^2}}} \, ,
\end{equation}
which allows us to estimate the order $k^2$ terms in the remaining
integral, leading to
\begin{equation}
  N_{\rm cyl} (k) = \frac{k^3 \left(\pi a^2L\right)}{6\pi^2} -\frac{k^2\left(2\pi aL+ 2\pi a^2\right)}{16\pi} +\ldots,
\end{equation}
in agreement with the first two terms in Eq.~(\ref{eq:Navge-four-terms}).


\begin{thebibliography}{99}
\bibitem{footnote2} The averaged
  asymptotic expression for $N(k)$ [see
  Eq.~(\ref{eq:Navge-four-terms})] is an expansion in powers of
  $1/k$, and  ignores fluctuations arising from the discrete nature
  of the eigenvalues.  In some cases these fluctuations are
  important as discussed in  Ref.~\onlinecite{balian-bloch-III}.
\bibitem{footnote3} Some authors assign
  a factor of 1/2 for the equality condition and 1 for
  the less than condition [see Eq.~(\ref{eq:Nkdefn-improved})].~\cite{baltes-hilf}
  This modification becomes important for
  some of the higher-order terms in the expansion to be discussed, and we will adopt this prescription when it is important.
\bibitem{footnote4} This expression is for bosons; an extra factor of two is usually  included when considering a degenerate Fermi
  gas.\cite{schroeder}
\bibitem{footnote5} See, for example,
  Refs.~\onlinecite{gutierrez-yanez, balian-bloch, waechter}.   
  Results related to Eq.~(\ref{eq:Navge-four-terms}) are sometimes
  given in terms of the density of states (rather than  the
  cumulative state number).  Also, many of the formal results related
  to the distribution of eigenvalues of Eq.~(\ref{eq:wave-equation})
  are  derived by solving related problems, such as the heat
  diffusion equation.  See Ref.~\onlinecite{balian-bloch} for a 
   discussion of this topic.
\bibitem{footnote7} For a   complementary
  description, we refer  interested readers to
  Ref.~\onlinecite{mulhall-moelter}.
\bibitem{footnote8} We
  can also use embedded {\tt Do} loops;  it is 
   tricky to determine the upper limits for each of the loops.
\bibitem{footnote9} It is possible to miss degeneracies due to
  rounding or other issues, which can be checked after the fact by
  comparing subsequent $\kappa$ values to determine if they are very close
  to each other.
\bibitem{footnote10} More mathematically sophisticated
  approaches, which correctly account for the averaging procedure, are
  described in Refs.~\onlinecite{baltes-hilf, gutierrez-yanez, balian-bloch,
    waechter}.  We agree with Lambert that, ``As an
  alternative to the rather inaccessible general proof, it seems
  worthwhile to give explicit proofs for the sphere and the cylinder
  since these shapes\ldots will tend to make the general result more
  plausible.''\cite{lambert} Reference~\onlinecite{balian-bloch} gives a
  beautiful introduction to a method involving the diffusion equation,
  Green's functions and a Laplace transform.
\bibitem{footnote12} We have used
  Eqs.~(\ref{eq:bound}) and (\ref{eq:alpha}) to determine $\chi$ near
  the boundary.  The largest deviation occurs for $n=1$, for which
  $\chi \sim 0.965\times\pi/4$ for large values of $\beta_{\ell,1}$ (a
  result that can be derived using Eq.~(\ref{eq:betal1})).
\bibitem{footnote13} One might wonder whether a correction is
  required for the points with $\ell=0$.  These points form a  line
  in $n$-$m$-$\ell$ space, so the corresponding correction would
  be of order $k$.  Note that we have inserted the degeneracy
  factor of $2\ell+1$ as a  sum in the first line of
  Eq.~(\ref{eq:Nvolsphere}), rather than as an integral.  If  we had done
  the integral from $-\ell$ to $\ell$ instead, we would have obtained
 $2\ell$ instead of $2\ell+1$, which would have yielded the same
  result at order $k^3$, but would have differed at order $k^2$.  The
  reason for the difference is that the bounding surfaces suspended
  above the $n$-$m$ plane (see Fig.~\ref{fig:nml}) contain the points
  with $m=\pm \ell$, and these points all lie  precisely on these
  surfaces.  If we had  done the integral in the $m$ direction, we would
  only have captured half of the points on these surfaces (see the
  discussion between Eqs.~(\ref{eq:A2ij}) and (\ref{eq:kappadefn}) in
  Sec.~\ref{sec:rectangular-parallelepiped}).  To correctly capture
  all of these points, we can  extend the range of the
  integration to be between $-\ell-1/2$ to $\ell+1/2$, which  
  gives the same result as the sum.
\end{thebibliography}
\end{document}